\documentclass[12pt,a4paper]{article}
\pdfoutput=1

\usepackage{soul}        % for spaced-out typesetting

\usepackage{amsmath,amsthm,amssymb,euscript,epsf,epsfig}
\usepackage{array}
\usepackage{fancybox}

\usepackage[usenames]{color}
\usepackage{amsfonts,bm}
\usepackage{graphicx}
\usepackage{enumerate}

\usepackage{parskip}     % For a blank line between paragraphs & no indenting
\usepackage[
      colorlinks=true,
      linkcolor=blue,
      urlcolor=blue,
      filecolor=blue,
      citecolor=blue,
      pdfstartview=FitH,
      pdftitle={},
      pdfauthor={},
      pdfsubject={},
      pdfkeywords={},
      pdfpagemode=UseNone,
      bookmarksopen=true
      ]{hyperref}
\usepackage[all]{hypcap}     %should be loaded AFTER hyperref, which should otherwise be loaded last.

\mathchardef\mhyphen="2D

\numberwithin{equation}{section}

\newcommand{\be}{\begin{equation}}
\newcommand{\ee}{\end{equation}}
\newcommand{\bea}{\begin{eqnarray}\displaystyle}
\newcommand{\eea}{\end{eqnarray}}

\newcommand{\nn}{\nonumber}

\def\eq#1{(\ref{#1})}

\def\beq{\begin{equation}}
\def\eeq{\end{equation}}
\def\beqa{\begin{eqnarray}}
\def\eeqa{\end{eqnarray}}
\def\bet{\begin{tabular}}
\def\eet{\end{tabular}}
\def\bs{\begin{split}}
\def\es{\end{split}}

\def\a{\alpha}

\def\G{\Gamma}
\def\e{\epsilon}

\def\m{\mu}

\def\n{\nu}

\def\k{\kappa}

\def\s{\sigma}
\def\t{\tau}

\def\cA{{\cal A}}  
  
\def\cG{{\cal G}}  
\def\cJ{{\cal J}}  
\def\cM{{\cal M}}  
  \def\cR{{\cal R}}

\def\mR{\mathbb{R}}

\newcommand{\rmx}{\mathrm{x}}

\def\one{{\hbox{\kern+.5mm 1\kern-.8mm l}}}
\def\zero{{\hbox{0\kern-1.5mm 0}}}

\definecolor{orange}{rgb}{1,0.5,0}

\newcommand{\ket}[1]{{\,| {#1} \rangle}}

\def\d{ \partial } 
\def\zb{{\bar z}}

\newcommand{\T}[3]{\ensuremath{ #1{}^{#2}_{\phantom{#2} \! #3}}}		%general tensor with upper indices displayed first 

\def\ha{\frac{1}{2}}

\def\pt{\widetilde{\psi}}

\def\d{\partial}
\def\db{\bar\partial}

\def\dbar{\ov\partial}

\def\ov{\overline}

\newcommand{\ex}[1]{{\rm e}^{#1}} 
\def\ii{{i}}

\newcommand{\ap}{\ensuremath{\alpha'}}

\newcommand{\bean}{\begin{eqnarray*}}
\newcommand{\eean}{\end{eqnarray*}}

\newcommand\dott[2]{#1 \! \cdot \! #2}

\def\eo{\overline{e}}

\begin{document}

\hfill \hbox{DFPD-11-TH-14} 

\vspace{-0.5cm}
\hfill \hbox{QMUL-PH-11-12}

\vspace{1cm}

\centerline{
\LARGE{ \textsc{New D1-D5-P geometries }} }

\vspace{0.5cm}

\centerline{
\LARGE{ \textsc{from string amplitudes}}}

\vspace{1cm}

\centerline{    
  \textsc{ Stefano Giusto$^{1,2}$, ~Rodolfo Russo$^3$, ~David Turton$^4$}  }

\vspace{0.5cm}

\begin{center}
$^1\,${Dipartimento di Fisica ``Galileo Galilei'',\\
Universit\`a di Padova,\\ Via Marzolo 8, 35131 Padova, Italy\\
}
\end{center}

\vspace{0.1cm}

\begin{center}
$^2\,${INFN, Sezione di Padova,\\
Via Marzolo 8, 35131, Padova, Italy}
\end{center}
\vspace{0.1cm}

\begin{center}
$^3\,${Queen Mary University of London,\\
Centre for Research in String Theory, School of Physics\\
Mile End Road, London E1 4NS, UK\\
}
\end{center}

\vspace{0.1cm}

\begin{center}
$^4\,${Department of Physics,\\ The Ohio State University,\\ Columbus, OH 43210, USA}
\end{center}

\vspace{0.2cm}

\begin{center}
{\small stefano.giusto@pd.infn.it, ~~r.russo@qmul.ac.uk, ~~turton.7@osu.edu}
\end{center}

\vspace{1cm}

\centerline{ 
 \textsc{ Abstract}}

\vspace{0.2cm}

{\small
We derive the long range supergravity fields sourced by a D1-D5-P
bound state from disk amplitudes for massless closed string emission.
We suggest that since the parameter controlling the string
perturbation expansion for this calculation decreases with distance from the bound state,
the resulting asymptotic fields are valid even in the regime of parameters 
in which there is a classical black hole solution with the same charges. The
supergravity fields differ from the black hole
solution by multipole moments and are more
general than those contained within known classes of solutions in
the literature, whilst still preserving four supersymmetries.  Our results
support the conjecture that the black hole solution should be
interpreted as a coarse-grained description rather than an exact
description of the gravitational field sourced by D1-D5-P bound states in this regime of parameters.
}
	
\thispagestyle{empty}

\vfill
\eject

\section{Introduction}

Type IIB string theory compactified on $S^1\times \cM_4$ (where
$\cM_4$ can be either $T^4$ or $K_3$) contains a large degeneracy of
configurations preserving 4 of the 32 supercharges of the trivial
vacuum. Since the seminal
papers~\cite{Strominger:1996sh,Callan:1996dv} we know that at zero
string coupling $g_s=0$, the number of these configurations matches
the Bekenstein-Hawking entropy of the extremal three-charge black
hole. It is expected that a non-renormalization theorem protects this
degeneracy from corrections as $g_s$ is switched on, explaining the
agreement between the number of string/D-brane configurations at zero
coupling and the entropy calculated in the black hole description. In
this gravitational regime, $g_s$ is non-zero but can be as small as we
like as long as the supergravity charges are large; for instance, this
requires that $g_s N>1$, where $N$ is any one of the D-brane charges of the configuration.

The microscopic derivation of the Bekenstein-Hawking entropy made this
class of extremal black holes an ideal arena for trying to address
other crucial questions at the heart of black hole physics. In
particular, a line of research advocated by Mathur and collaborators
has focused on the problem of studying the geometrical backreaction of
individual microscopic D-brane configurations (microstates) as the parameter
$g_s N$ is increased from zero to a large value (see the reviews~\cite{Mathur:2005zp,Mathur:2008nj,Skenderis:2008qn,Balasubramanian:2008da,Chowdhury:2010ct}). 
The central question is to understand whether the different
elementary configurations produce distinguishable gravitational
backgrounds and, if so, to determine the scale at which the differences start
being relevant. 

This question is closely tied with the information paradox:
the Hawking emission process resulting from a classical black hole metric coupled to quantum fields leads to a breakdown of unitarity~\cite{Hawking:1974sw,Hawking:1976ra} so if the physical black holes we observe in Nature are to be accurately 
described by quantum mechanics, then one requires a more refined description of physics at the horizon than that provided by Hawking's description (for recent progress in this area see~\cite{Mathur:2009hf,Mathur:2011wg,Mathur:2011uj}). 
One way to avoid the conclusions of the Hawking theorem is to seek a more refined description of the gravitational field sourced by physical black holes;
one of the crucial features of the fuzzball proposal~\cite{Mathur:2005zp,Mathur:2008nj} is that sizable deviations 
from the `naive' black hole geometry appear at a scale proportional to $g_s N$.

One way to test this proposal is to look for solutions of type IIB
supergravity which preserve four supersymmetry generators and have the
same D1, D5 and Kaluza-Klein charges of the usual black hole, but which
differ from the `naive' black hole geometry at a large scale
$\sim g_s N$ (for a review, see~\cite{Bena:2007kg}). This program builds on the successful analysis of the
two-charge D1-D5 configurations, in which case the Kaluza-Klein
momentum charge is set to zero. By dualizing the supergravity solutions for a fundamental string with a travelling wave~\cite{Dabholkar:1995nc,Callan:1995hn}, a large class of horizon-less
geometries was found and studied~\cite{Lunin:2001fv,Lunin:2001jy} which led to the fuzzball proposal and many further studies, see e.g.~\cite{Taylor:2005db,Kanitscheider:2007wq}. In the D1-D5 duality frame these solutions are everywhere smooth and free of brane sources, a feature which is however duality-frame dependent. 

The analysis of the three-charge D1-D5-P configurations has proved much more
challenging. A large class of $1/8$-BPS smooth and horizon-less supergravity solutions is known~\cite{Bena:2004de,Bena:2005va,Berglund:2005vb,Bena:2006is,Bena:2007qc,Balasubramanian:2006gi,deBoer:2008zn,Bena:2010gg}, 
however quantizing the degeneracy of these geometries does not yield an entropy of the same order as the entropy of the
black hole~\cite{deBoer:2009un}. Another complication is that the precise relation between the known
supergravity solutions and the microstates is less clear than in the
two-charge case~\cite{Skenderis:2007yb}. One way to match the geometrical and the microscopic
descriptions is to use the AdS/CFT dictionary; see for
instance~\cite{Giusto:2004id,Giusto:2004kj,Ford:2006yb} for an explicit implementation of this
approach (see also \cite{Bena:2011zw} for recent progress in understanding 
the AdS/CFT dictionary in different phases of this system). 
The basic idea is the following: if a solution is indeed
dual to the elementary states of the extremal three-charge black hole,
then it should have a ``near-horizon'' limit, where its asymptotic
geometry is AdS$_3 \times S^3 \times \mathcal{M}_4$; in this limit one can use
the AdS/CFT correspondence to read from the geometrical data the
corresponding state in the CFT description. However it is more
difficult to reverse the logic and use this approach to {\em construct}
new supergravity solutions starting from the definition of a CFT microstate.

In this paper, we use a different approach to derive the
large-distance backreacted geometry from a microscopic configuration
of D-branes. We exploit the fundamental
definition of D-branes as space-time defects which introduce borders
in the string world-sheet and identify the left and the right moving
excitations of closed strings. By calculating string amplitudes
describing the emission of each massless closed string field from
world-sheets with disk topology one can derive the large distance
fall-off of the various supergravity fields sourced by a given D-brane
bound state. This was originally done for the simplest case of
$1/2$-BPS configurations~\cite{DiVecchia:1997pr,DiVecchia:1999uf}. Of
course, the leading long-range behaviour of a solution is determined
by its charges and, if one repeats the same calculation for a simple
superposition of D1 and D5-branes, then only the `naive' D1/D5
supergravity solution at large distances is reproduced and, as
expected, no higher multipole moments appear.

The backgrounds we are interested in are however more complicated and
should correspond to states in the Higgs branch of the D1/D5
world-volume CFT.  While a precise quantum description of the states
in the Higgs branch is difficult, semiclassically we can characterize
them by giving a non-zero vacuum expectation value (vev) to the
massless fields in the spectrum of the open string 
stretched between the D1 and the D5-branes. Even
this is quite challenging, as, at the CFT level, the vertex operators
for the open strings stretched between different D-branes contain
complicated boundary changing operators known as twist/spin fields, as
we describe in Section~\ref{tos}.  However the large-distance
expansion of the corresponding backreacted geometry corresponds {\em
  order by order} to the standard open string perturbative expansion,
where also the open string vevs are treated perturbatively, something
which can be done quite straightforwardly as long as we do not have to
deal with too many twist fields. Recently it was
shown~\cite{Giusto:2009qq} that this approach captures precisely the
first non-trivial moments of the two-charge geometries in the D1/D5
duality frame.

The reason for this correspondence lies in how the superghost charge
is saturated in superstring perturbation theory. As summarized in
Section~\ref{sec:backreaction}, the main space-time implication of the
world-sheet constraint following from the superghost correlators is
that the expansion parameter in the perturbative calculation of the
geometry produced by each D-brane configuration is $g_s N
\alpha'/r^2$, where $r$ is the radial distance in $\mathbb{R}^4$ at which we
are probing the background. So at large enough distances the results
derived from string amplitudes should be reliable also in the black
hole regime, where $g_s N$ and the Higgs vevs are large. 

In this paper we combine the D1/D5 setup of~\cite{Giusto:2009qq} with
the microscopic description of a null-wave on
D-branes~\cite{Das:1996ug,Hikida:2003bq,Blum:2003if,Bachas:2003sj} to
provide a description of the three-charge microstate at least at the
semiclassical level. It was shown in~\cite{Black:2010uq} that the
boundary state for a D-brane with a travelling wave can be used to
derive the large distance behaviour of the two-charge configuration in
the D-brane/momentum duality frame. Thus, at the microscopic level,
the basic building block for the most general three-charge
configuration is provided by a set of D1 and D5-branes, each one with
an a priori different wave profile, and a non-trivial vev for the open
strings stretched between the two types of D-branes.

We find that the world-sheet analysis of this generic building block
is rather non-standard, as the sector of the D1/D5 open strings is
described by a logarithmic CFT~\cite{Gurarie:1993xq} (for related work see e.g.~\cite{Periwal:1996pw,Kogan:1996zv,Kogan:2000fa,Lambert:2003zr}). 
We focus our attention in this paper to the simplest case where the wave profiles on the D1 and
D5-branes are identical and the worldsheet description is given in
terms of a standard CFT\footnote{This class of microstates was
recently studied at the level of the probe-brane approximation
in~\cite{Bena:2011uw}.}.

We calculate for these configurations the one-point functions of the
massless supergravity fields with disk topologies, and are
particularly interested in the contributions which vanish in all
two-charge limits, which we describe as `{\em new}' contributions.  We
derive the new part of the large-distance expansion of the backreacted
geometry up to order $1/r^4$.  There are other terms in the $1/r$
expansion of the solution that follow from the non-linearity of
gravity, rather than the non-trivial structure of the microstates; in
our string approach these contributions are related to world-sheet
surfaces with more than one border. We will not try to derive these
terms from string amplitudes, as it is much easier to do so by solving
the standard supergravity equations and by using the results from the
disk amplitudes as boundary conditions.

Using the type IIB supersymmetry equations we explicitly check that
the supergravity backgrounds derived from these string amplitudes
preserve four supercharges.  A surprising result is that we find
supergravity fields not contained within known classes of $1/8$-BPS
solutions in the literature~\cite{Bena:2007kg}: in particular the
NS-NS 2-form and the R-R 0 and 4-form potentials are non-trivial,
while the metric of the $\mathbb{R}^4$ part is still hyper-Kahler, apart from
the presence of a warp factor.

Our calculation supports the conjecture that  
individual states should have backreactions with non-trivial 
multipole moments, whilst an appropriate thermodynamic ensemble average 
would average out these moments to zero, obtaining the `unique' classical black hole solution with horizon~\cite{Mathur:2005zp,Balasubramanian:2005mg,Balasubramanian:2006jt,Balasubramanian:2007zt}.

The paper is structured as follows. In Section~\ref{sec:ansatz}, we
analyze the type IIB supersymmetry equations perturbatively in $1/r$ 
to constrain the form of $1/8$-BPS geometries. We focus
on configurations in which the structure space $\cM_4$ only receives
an overall warping from the microstate backreaction. We obtain from
supergravity a set of conditions which we later use to test the
consistency of our string derivation.  In
Section~\ref{sec:Worldsheet_tech}, we review the basic ingredients
necessary for the string computation and show why the case of
identical profiles on the D1 and D5 branes has a standard CFT
description. In Section~\ref{sec:backreaction} we calculate the disk
one-point functions for the massless supergravity fields and in
Section \ref{sec:comparison} we derive from this data the
large-distance behaviour of the supergravity background corresponding
to each microscopic D-brane configuration.  Finally, in
Section~\ref{sec:discussion}, we present our conclusions discussing
limitations and possible generalizations of this approach for
extracting information about the backreactions of D-brane microstates
from string amplitudes. The appendices contain technical details
of the type IIB supersymmetry equations and the string vertices used
in the main text.

\section{Supersymmetry analysis} \label{sec:ansatz}

The geometry sourced by a D1-D5-P bound state must preserve the same four
supersymmetries as the `naive' black hole with the same charges. We
derive in this section, from the supergravity point of view, the
constraints imposed on the geometry by the existence of these four conserved
supercharges. Though we do not input supersymmetry explicitly in
computing string amplitudes, the string results should be compatible
with these supersymmetry constraints. This will thus provide a useful
and non-trivial check on the string amplitude computation.  For the
purpose of comparison with the world-sheet results it is sufficient to
restrict the supergravity analysis to $1/r^4$ order in the asymptotic
expansion. We have checked that the analysis can be extended to all
orders in $1/r$, but we will leave the details of the exact solution
for a future work.

\subsection{The ansatz}

We consider a general ansatz for a IIB configuration compactified on
$T^4 \times S^1$ which does not break the isometries along the $T^4$ directions.
For the NSNS fields, a generic ansatz with this property is (in the string frame):
\bea
ds^2 &=& \frac{1}{\sqrt{Z_1 Z_2}}\Bigl[-\frac{1}{Z_3}\, d{\hat t}^2 + Z_3\, d{\hat y}^2 \Bigr]+\sqrt{Z_1 Z_2}\, ds^2_4+\sqrt{\frac{Z_1}{Z_2}}\, ds^2_{T^4} \,, \cr
B &=& -b_0\,d\hat t \wedge d\hat y + b_1\wedge d\hat y + \widetilde b_1\wedge d\hat t + b_2\,, \cr
e^{2\phi} &=& D\,, \label{NSNSsusy}
\eea
where we have introduced the short-hand notation
\be
d\hat t = dt+k \,,\quad d\hat y = dy+dt-\frac{dt+k}{Z_3}+a_3\,.\label{dtdy}
\ee
Here $ds^2_4$ is a generic Euclidean metric on $\mathbb{R}^4$, $ds^2_{T^4}$ is the flat metric on $T^4$ (which we take to be $ds^2_{T^4} = \delta_{ab}\,dz^a dz^b$), $Z_1$, $Z_2$, $Z_3$, $b_0$, $D$ are 0-forms, $k, a_3, b_1, \widetilde b_1$ 1-forms, and $b_2$ a 2-form on $\mathbb{R}^4$.
Similarly we can write for the RR 0-form and 2-form
\bea
C^{(0)} &=& c \,, \cr
C^{(2)} &=& -\frac{1}{\widetilde Z_1} d\hat t \wedge d\hat y + a_1\wedge d\hat y + \widetilde a_1\wedge d\hat t + \widetilde \gamma_2\,,\label{RRsusy}
\eea
with $c, \widetilde Z_1$ 0-forms, $a_1, \widetilde a_1$ 1-forms and $\widetilde \gamma_2$ a 2-form on $\mathbb{R}^4$. The RR 4-form is constrained to have a self-dual field strength $F^{(5)}=* F^{(5)}$, hence one can write
\bea
F^{(5)} &=& f_1 \,d\hat t \wedge dz^4 + f_2 \,d\hat y \wedge dz^4  + g \wedge dz^4 \cr
	&& {}+Z_2^2 Z_3\,f_1 \,d\hat y \wedge dx^4+\frac{Z_2^2}{Z_3}\,f_2 \,d\hat t \wedge dx^4 +\frac{Z_2}{Z_1}\, *_4 g\wedge d\hat t \wedge d\hat y\,,
\eea
where $f_1, f_2$  are 0-forms,  $g$ a 1-form on $\mathbb{R}^4$, and $dz^4$ and $dx^4$ are the volume forms of $T^4$ and $\mathbb{R}^4$.

At order $1/r^2$ the solution should reduce to the `naive' D1-D5-P black hole; we will moreover use some foresight from the string computation to assert that the 1-forms $k, a_1, \widetilde a_1, a_3, b_1, \widetilde b_1$ receive non-trivial contributions first at order $1/r^3$ and that the quantities
$c, f_1, f_2, g, b_0, b_2$ and the metric $ds^2_4$ do not have non-trivial terms until order $1/r^4$. 

Under these assumptions, and discarding terms of order higher than $1/r^4$, the equations of motion for $B$ and $C^{(2)}$ can be approximated by $d * B=0$, $d  * C^{(2)}=0$, and imply, in particular, that
\be
d b_2 = *_4 d\widetilde b_0\,,\quad d\widetilde \gamma_2 = *_4  d \widetilde Z_2\,,\label{deltagamma}
\ee
for some 0-forms $\widetilde b_0$ and $\widetilde Z_2$. Moreover the Bianchi identity $dF^{(5)}=0$ implies
$df_1=df_2=0$, so that one has $f_1=f_2=0$, and $dg=d *_4 g=0$, so that $g = d f$ and $d *_4 d f=0$, for some 0-form $f$. Then $F^{(5)}$ can be simplified to 
\be
F^{(5)} = d f \wedge dz^4+ *_4 df \wedge d\hat t \wedge d\hat y\,.\label{f5susy}
\ee
We also know that in the `naive' black hole geometry $Z_1 = \widetilde Z_1$, $Z_2 = \widetilde Z_2$, $D=Z_1/Z_2$, and we allow these identities to be modified at order $1/r^4$.

In summary our ansatz is given by (\ref{NSNSsusy}), (\ref{dtdy}), (\ref{RRsusy}), (\ref{deltagamma}), (\ref{f5susy}), with the asymptotic boundary conditions
\bea\label{asymptoticbc}
&& Z_1,\, Z_2, \,Z_3 = 1+ O(r^{-2})\,,\quad \widetilde Z_i = Z_i + O(r^{-4})\quad (i=1,2)\,,\quad D = \frac{Z_1}{Z_2}+O(r^{-4})\,,\cr
&& b_0,\, \widetilde b_0, \,c,\,f = O(r^{-4}) \,,\quad ds^2_4 = dx_i dx_i + O(r^{-4})\,,\quad k,\, a_1,\, \widetilde a_1,\, a_3,\, b_1, \widetilde b_1 = O(r^{-3})\,.\nonumber\\
\eea

\subsection{Results}
It is a straightforward though quite lengthy exercise to impose the vanishing of the dilatino and gravitino supersymmetry variations and derive the condition this imposes on the various metric coefficients\footnote{Though there are many papers that study the supersymmetry conditions for 5D minimal supergravity coupled to various vector and hypermultiplets, as far as we understand none of those results directly apply to the case specified by our ansatz.}. Some details of this computation, up to order $1/r^4$, are given in Appendix \ref{sec:susy_details}. The results are 
\bea
&&\widetilde Z_1= Z_1\,,\quad \widetilde Z_2 = Z_2\,,\quad D = \frac{Z_1}{Z_2}\,,\quad b_0 =\widetilde b_0= c = f  \,,\cr
&& \widetilde a_1=a_1\,,\quad \widetilde b_1 = b_1\,,\quad da_1 = *_4 d a_1\,,\quad da_3 = *_4 d a_3\,,\quad db_1 = *_4 d b_1\,,\cr 
&&\overline{R}^{(4)}_{ij,kl} = \frac{1}{2}\,\epsilon_{ijrs}\,\overline{R}^{(4)}_{rs,kl}\,,\label{susyconditions}
\eea 
where $\overline{R}^{(4)}_{ij,kl}$ is the Riemann tensor of $ds^2_4$. The last condition is equivalent to require that $ds^2_4$ be hyper-Kahler. 

Moreover the gauge field equations of motion and the $ty$ component of Einstein's equations  imply, at this order, that
\be
d *_4 d Z_1 =0\,,\quad d *_4 d Z_2 =0\,,\quad  d *_4 d Z_3 =0\,,\quad d *_4 d b_0 =0\,,\quad d *_4 d k =0\,.\label{eomconditions}
\ee
Conditions (\ref{susyconditions}) and (\ref{eomconditions}) are enough to guarantee that all other components of Einstein's equations be satisfied. 

\section{String world-sheet setup} \label{sec:Worldsheet_tech}

\subsection{D-brane configuration}

We consider type IIB string theory on $\mathbb{R}^{1,4}\times
S^1\times T^4$. We denote the 10D coordinates $(x^{\m},\psi^{\m})$
by $ \m, \n =t,y,1,\ldots,8$. We use $(i,j,\ldots)$ and $x^1, \ldots , x^4$ for the $\mathbb{R}^{4}$ directions, we use $(a,b,\ldots) \,$ and $x^5, \ldots , x^8$ for the $T^4$ directions
and we use $(I,J,\ldots)$ to refer to the combined $\mathbb{R}^{1,4}\times S^1$ directions.
We work in the light-cone coordinates
\be
v = \left( t + y \right), \qquad  u = \left( t - y \right) ~
\ee
constructed from the time and $S^1$ directions.
In Appendix~\ref{app:vertices} we collect our CFT conventions, including the form of the BRST charge and some details on the closed string vertices, while in Appendix \ref{app:conventions} we record our conventions for spinors.

We focus on $1/8$-BPS configurations composed of D1 and D5-branes with a
non-trivial wave. The branes have common Neumann (Dirichlet)
boundary conditions along the directions $t$, $y$ ($x^i$ in the
$\mathbb{R}^4$), while they have mixed Neumann/Dirichlet boundary conditions in
the $T^4$. We can summarize the D-brane configuration under study in
the following table:
\be
\begin{array}{c|cc|c|c}
   &   v  &   u  & \mR^4  &  T^4   \\ 
D1 & \rmx & \rmx & f^{\rm D1}_i(v) & f^{\rm D1}_a(v)  \\
D5 & \rmx & \rmx & f^{\rm D5}_i(v)  & \rmx  \\
\end{array}
\ee
where ``x'' denotes a Neumann direction and $f$ indicates the
($v$-dependent) position of the D-brane in the Dirichlet directions.
We will ignore from the very beginning the profile along the $T^4$ by
setting $ f^{\rm D1}_a = 0$.  Initially we will allow for
independent wave profiles $f^{\rm D1}$ and $f^{\rm D5}$, before focusing our calculations on the case in which the two
profiles are identical.

\subsection{Boundary conditions for 1-1 and 5-5 strings}

We now review the boundary conditions for an open string with both
endpoints on a D-brane carrying a travelling wave.  Encoding the
effect of the D-brane profile has the effect of resumming all the open
string insertions of the vertex
\be \label{Vf}
V_f = \int \left(\frac{1}{2\alpha'} f_j \, \partial X^j + {\dot f}_j \psi^j \psi^v \right) dz
\ee
describing the KK charge of the D-brane configurations (see
e.g.~\cite{Callan:1988wz,Bachas:2002jg}) so our results will be exact
in this respect.

The boundary conditions on the worldsheet fields in the open string picture may be expressed in terms of a reflection matrix $R$ as 
\bea
\pt^\m &=& \eta \, \T{R}{\m}{\n}(V) \psi^\n    \label{psii} \\
\dbar X^\m &=&  \T{R}{\m}{\n}(V) \d X^\n - \T{\delta}{\m}{u} \, 8
\ap \ddot{f}_j \psi^j \psi^v 
\label{nln}
\eea
where we use a capital $V$ to indicate the string field corresponding
to the coordinate $v$. The parameter $\eta$ can be set to 1 at $\s=0$,
while at $\s=\pi$ we have $\eta = 1$ or $\eta = -1$ corresponding to
the NS and R sectors respectively.

For 1-1 strings, the holomorphic and the anti-holomorphic world-sheet
fields are identified with the reflection matrix $R = R_{\rm D1}$
where (see~\cite{Black:2010uq} and references within)
\be \label{eq:D1_bcs}
\big( R_{\rm D1} \big)^{\m}_{\phantom{\m \!}\n} ~=~ 
\left( \begin{array}{cccc} 
        	  1              &      0      &          0           &          0          \\
   	  4 |\dot{f^{\rm D1}}({V})|^2   &      1      &  -4 \dot{f}^{\rm D1}_i({V})   &          0          \\
 	    2 \dot{f}^{\rm D1}_i({V})   &      0      &    \!\! - \one       &          0          \\
  	        0              &      0      &          0           &   \!\! - \one
\end{array} \right) \;,
\ee 
where $\one$ denotes the four-dimensional unit matrix and the indices follow the ordering $(v,u,i,a)$. 

Similarly, for 5-5 strings ending on a D5-brane with profile $f^{\rm D5}$, the right
and left-moving world-sheet fields are identified with $R_{\rm D5}$,
where
\be \label{eq:D5_bcs}
\big( R_{\rm D5} \big)^{\m}_{\phantom{\m \!}\n} ~=~ 
\left( \begin{array}{cccc} 
        	  1              &      0      &          0           &          0          \\
   	  4 |\dot{f}^{\rm D5}({V})|^2   &      1      &  -4 \dot{f}^{\rm D5}_i({V})   &          0          \\
 	    2 \dot{f}^{\rm D5}_i({V})   &      0      &    \!\! - \one       &          0          \\
  	        0              &      0      &          0           &     \! \one
\end{array} \right) .
\ee
We note that the reflection matrices $R_{\rm D1}$ and $R_{\rm D5}$
preserve the Minkowski metric
\be \label{eq:R_preserves_eta}
R^{T}_{\rm D1\,,D5} \, \eta \, R_{\rm D1\,,D5} = \eta~.
\ee
Importantly, this setup differs from the case of D-branes at angles or with a constant magnetic field
in that our reflection matrices contain a non-trivial function of the coordinate $V$. 
Thus, when we use the identification~\eqref{psii} on the antiholomorphic part
$\widetilde{T}_{\widetilde{\psi}}$ of the fermionic stress energy tensor
(see~\eqref{Te} for our conventions), we obtain a new term involving
$\ddot{f}$. In the full stress energy tensor, this is cancelled by a
similar term in $\widetilde{T}_{X}$ coming from the non-linear part
of the bosonic identification~\eqref{nln}. So all terms involving
$\ddot{f}$ cancel and, thanks to~\eqref{eq:R_preserves_eta}, we find
that our boundary conditions define the usual open string Virasoro
algebra.

\subsection{Boundary conditions for 1-5 and 5-1 strings}

For a string with one endpoint on a D1-brane and one endpoint on a
D5-brane, the situation is more complicated. Boundary conditions for
a string with endpoints on different D-branes are discussed
in~\cite{Bertolini:2005qh}, from which we now review some relevant
expressions. Denoting the bosonic string coordinates by $x^{\mu}$, the
boundary conditions at the two endpoints of the string may be written
as
\begin{equation}
\db x^{\mu}\Big|_{\sigma=0,\pi} ~=~ 
\big( R_\sigma \big)^{\m}_{\phantom{\m \!}\n}\,  \d x^{\nu}\Big|_{\sigma=0,\pi}~. 
\label{bc1}
\end{equation}
For a string with both endpoints on the same D-brane, we have $R_0 = R_{\pi}$ and one may solve the boundary conditions by writing $x^{\m}$ in terms of a holomorphic field $X^{\mu}(z)$:
\begin{equation} 
x^{\mu}(z,\zb) = q^{\mu}+ \ha \Big[X^{\mu}(z) +\big(R_0\big)^{\m}_{\phantom{\m \!}\n} \,X^{\nu}(\zb)\Big]~.
\label{sol1}
\end{equation}
For a string with endpoints on different D-branes, we may define
multi-valued fields $X^{\mu}(z)$ and introduce a branch cut in the
$z$-plane just below the negative real axis, such that $\d X^{\mu}$
has a monodromy written in terms of a ``monodromy matrix'' $M$:
\begin{equation}
\d X^{\mu}({\rm e}^{2\pi i}z) ~=~ \T{M}{\m}{\n} \,\d X^{\nu}(z)
~,~~\mbox{where}~~~~M \equiv R_\pi^{-1} R_0~.
\label{eq:monodromy}
\end{equation}
Then the boundary conditions are again solved by~\eqref{sol1}, but now
with a multivalued field $X^{\m}(z)$. 
In our case we have 
\be \label{eq:D1-D5_bcs}
R_{0} ~=~ R_{\rm D1}\,, \qquad R_{\pi} ~=~ R_{\rm D5}  \,.
\ee
Then the monodromy matrix turns out to be only a function of the difference between the two profiles, and has the same form as the monodromy matrix for one profile (with the extra minus sign in the $T^4$ directions):
\be \label{Mf}
\T{M}{\m}{\n} ~=~  \big( R_{\pi}^{-1} R_0 \big)^{\m}_{\phantom{\m \!}\n} ~=~
\left( \begin{array}{cccc} 
        	  1              &      0      &          0          &         0         \\
   	  4 \, |\dot{f}^{\rm D5}-\dot{f}^{\rm D1}|^2   &      1      &   4 (\dot{f}^{\rm D5}-\dot{f}^{\rm D1})_i  &   	   0         \\
 	    2 (\dot{f}^{\rm D5}-\dot{f}^{\rm D1})_i   &      0      &    \!  \one         &         0         \\
  	        0              &      0      &          0          &   \!\! - \one
\end{array} \right) .
\ee 
The monodromy matrix has a similar form to that studied in the context
of null orbifolds~\cite{Liu:2002ft,Liu:2002kb}. This can be seen by
defining
\be
f^-_i ~=~ f^{\rm D5}_i-f^{\rm D1}_i~,
\ee
and then by writing 
\be
M ~=~ M_0 \exp \left( 2\pi \dot{f}^-_i \cJ_i \right)~,
\ee
where $M_0$ is the monodromy matrix for $f^-=0$ (see~\eqref{M0m}) and
\be \label{eq:fdot_dot_J}
\T{(\dott{\dot f^-}{\cJ})}{\m}{\n} ~=~  \frac{1}{\pi}
\left( \begin{array}{cccc} 
      0     & 0 &      0     &  0 \\ 
      0     & 0 & 2 \dot f^-_i &  0 \\ 
  \dot f^-_i  & 0 &      0     &  0 \\
     0      & 0 &      0     &  0   
\end{array} \right) ~. \quad
\ee
Thus the monodromy~\eqref{eq:monodromy} for the coordinates  $\d X$
can be written as follows 
\be
\d X^{\mu}({\rm e}^{2\pi i}z) ~=~ \left[ \exp \left( 2\pi \dot{f}^-_j \cJ_j \right) \right]^{\m}_{~\n} \d X^{\nu}(z)~,
\ee
The unusual feature is that $\dot{f}^- \cJ$ is nilpotent and thus
not diagonalizable. A consequence of this is that the sector of open
strings stretched between D-branes with different profiles has the
structure of a logarithmic CFT. In particular, one finds that at each
level there is a Jordan block of rank three, related to the property
$(\dot{f}^- \cJ)^3=0$. We shall postpone the study of the full
analysis of this problem to later work, and in this paper we shall
treat the simpler scenario in which the profiles are equal, in which
case one has $f^- = 0$.

\subsection{Equal D1 and D5 profiles} \label{epro}

We now specialize to the case in which both the D1 and the D5 branes
are wrapped $n_w$ times around~$y$. Letting the length of the $y$
direction be $2 \pi R$, each brane then has total length $L_T = 2
\pi n_w R$ and we use $\hat{v}$ for the corresponding world-volume
coordinate on the D-branes, having periodicity $L_T$. Moreover we
focus on the case in which the D1 and D5 branes have identical
profiles, which we denote by $f \equiv f^{\rm D1} = f^{\rm D5}$.
Clearly this common profile $f^i$ satisfies 
$f^i(\hat{v}+L_T) = f^i(\hat{v})$.

We see from the discussion in the previous subsection that in this
case the monodromy matrix reduces to that of the two-charge D1-D5
system as studied in~\cite{Giusto:2009qq}, i.e.
\be\label{M0m}
\T{(M_0)}{\m}{\n} =
\left( \begin{array}{cccc} 
        1  &      0      &   0       &         0         \\
   	  0  &      1      &   0       &   	     0         \\
 	  0  &      0      &  \!  \one &         0         \\
  	  0  &      0      &   0       &   \!\! - \one
\end{array} \right) .
\ee 
In this case the worldsheet CFT is not logarithmic and one can use the
boundary conditions \eq{eq:D1_bcs} with $f^{\rm D1} = f$ in the
calculation of closed string emission from a D1-D5 disk. This is the
approach we follow in Section \ref{sec:2_boundaries}.

We will later also require the boundary conditions for the left and
right moving spin fields. These are
\be
\widetilde S^{\hat{A}} = (\cR_{\rm D1})^{\hat A}_{~\hat B} \, S^{\hat B} 
\ee
where $\cR_{\rm D1}$ is the spinor representation of the reflection
matrix $R_{\rm D1}$. For a flat D1 brane, $\cR_{\rm D1} = \G^{ty}$
however for a D-brane with a travelling wave we can read $\cR_{\rm
  D1}$ from the R-R zero mode boundary state for a D1 brane with
travelling wave described by the profile $f^i$. This was calculated in
\cite{Black:2010uq} to be
\be \label{eq:B_D1P}
|D1;P\rangle _{\psi,0}^{(\eta)} = {\cal M}_{AB}^{(\eta)}\,\ket{A}_{-\frac{1}{2}} \ket{\widetilde B}_{-\frac{3}{2}}
\ee
where, following our spinor conventions (given in Appendix \ref{app:conventions}) in which the gamma matrices in ten dimensions are denoted $\G_{(10)}^{\m}$, we have
\be \label{eq:M_D1P}
{\cal M}^{(\eta)} ~=~ i \, C \left( \frac{1}{2} \G_{(10)}^{uv} + \dot{f}^i(v)\G_{(10)}^{iv} \right)  \left( \frac{\one - i \eta \G_{11}}{1- i \eta} \right),
\ee
so we read off
\be \label{eq:cR_D1P}
\cR_{\rm D1} ~=~ \left( \frac{1}{2} \G_{(10)}^{uv} + \dot{f}^i(v)\G_{(10)}^{iv} \right).
\ee

\subsection{Twisted open string vertex operators} \label{tos}

In Section \ref{sec:2_boundaries} we calculate amplitudes on a mixed disk with half its boundary on a D1 and the other half on the D5, and two twisted vertex operator insertions as studied in~\cite{Billo:2002hm,Giusto:2009qq}, details of which we record here. 

Since the monodromy matrix is now given simply by \eq{M0m}, we use the same twisted open string vertices considered in \cite{Giusto:2009qq}, which take the form
\beq\label{Vop15} 
V_{\mu} = \mu^{A} \ex{-{\varphi\over 2}} S_A \,
 \Delta \,, \quad\qquad ~~~~~~~
V_{\bar \mu} =\bar \mu^{A} \ex{-{\varphi\over 2}}  S_A \, \Delta \,
\eeq
where $\mu^A$ and $\bar{\mu}^A$ are Chan-Paton matrices with
$n_1\times n_5$ and $n_5\times n_1$ components respectively, $S_A$ are
the $SO(1,5)$ spin fields, $\varphi$ the free boson appearing in the
bosonized language of the worldsheet superghost $(\beta,\gamma)$, and
$\Delta$ is the bosonic twist operator with conformal dimension
${1\over 4}$ which acts along the four mixed ND directions (in which the monodromy matrix has the value $-1$) 
and changes the boundary conditions from Neumann to Dirichlet and vice versa.

We focus on open string condensates involving
 only states from the Ramond sector.  Notice that states in the Ramond
 sector will break the $SO(4)$ symmetry of the DD directions
 $\mathbb{R}^4$, while they are invariant under the $SO(4)$ acting on
 the compact $T^4$ torus.
The most general condensate of Ramond open strings can be written as:
\be\label{mubarmu}
\bar \mu^A \, \mu^B=v_I (C\Gamma^I)^{[AB]}+
\frac{1}{3!} \,v_{IJK} (C\Gamma^{IJK})^{(AB)}~,
\ee
where the parenthesis on the indices $A, B$ are meant to remind that
the first term is automatically antisymmetric, while the second one is
symmetric. Thus the open string bispinor condensate is specified by a
one-form $v_I$ and an self-dual three-form $v_{IJK}$. The self-duality
of $v_{IJK}$ follows from $\bar\mu^A$ and $\mu^B$ having definite 6D
chirality and can be written as
\beq\label{sdv}
v_{IJK} = \frac{1}{3!} \epsilon_{IJKLMN} v^{LMN}\,.
\eeq
In this paper we shall consider only the components of $v_{IJK}$ which have one leg in the $t,y$ directions and two legs in the $\mR^4$; this choice of components was associated to considering profiles only in the $\mR^4$ directions in~\cite{Giusto:2009qq}.
In $t,y$ coordinates with $\e_{1234}=1$, we see that \eq{sdv} becomes
\be \label{vtij}
v_{yij} =  \ha \epsilon_{ijkl} v_{tkl}~.
\ee
The self-duality properties in light-cone coordinates are then
\be
v_{uij} = -\frac{1}{2}\epsilon_{ijkl}\,v_{ukl}\,,\qquad v_{vij} = +\frac{1}{2}\epsilon_{ijkl}\,v_{vkl}\,.
\ee
Since the spinors $\bar \mu^A$ and $\mu^B$ carry $n_5\times n_1$ and $n_1\times n_5$ Chan-Paton indices, the condensate $\bar\mu^{A}\mu^{B}$ must be thought of as the vev for the sum
\be
  \sum_{m=1}^{n_1} \sum_{n=1}^{n_5}  \bar\mu^{A}_{m n}\,  \mu^{B}_{n m}\,,
\ee
which, for generic choices of the Chan-Paton factors, is of order $n_1 n_5$.

\section{D-brane geometrical backreaction} \label{sec:backreaction}

In this section we derive the geometrical backreaction of the D-brane
configuration discussed in the previous section, starting from the
couplings of the closed string massless fields. In perturbation
theory, these couplings are captured by the one-point functions of each
closed string state on world-sheets with boundaries. As we are not
interested in purely quantum gravity effects ({\em i.e.}~couplings
weighted by the Planck length) we will ignore all contributions to the
one-point functions from topologies which have handles and focus
only on planar world-sheets with boundaries. The non-trivial vacuum
expectation values for the open string fields~\eqref{Vf} and~\eqref{Vop15} should ensure that we are considering, at least
semiclassically, not merely a naive superposition of three charges but
a real bound state.

As discussed in the previous section, the boundary
conditions~\eqref{psii} and~\eqref{nln} resum all the open string
insertions describing the KK momentum charge of the D-brane configurations, so
our results will be exact in this respect. On the contrary we will
treat perturbatively the open string insertions~\eqref{Vop15} related to
the vev of the strings stretched between the D1 and D5 branes.

The interesting microstates, for which we might expect a gravitational
description, have large open string vevs and so, in principle, we
should resum amplitudes with many twisted vertices.  However, as
briefly mentioned in the~Introduction, it is possible to check that
string amplitudes with a different number of open string
insertions~\eqref{Vop15} contribute to different terms in the large
distance expansion of the corresponding gravity solution.  The
argument goes as follows: disk amplitudes are non vanishing only if
the total superghost charge of the correlator is $-2$; the open string
vertices~\eqref{Vop15}, which must be always paired in order to have a
consistent boundary, are in the $-1/2$ picture, so each insertion of
$V_\mu$, $V_{\bar\mu}$ in a non-zero correlator requires an extra $e^{\phi}$
factor to keep the correlator non-trivial. 

Thus the expansion in the number of twisted open string insertions is actually weighted by the
$e^{\phi}$ charge we need to saturate. This can be done by inserting
in the amplitude the supercurrent, or equivalently by changing the
picture of the emitted closed string vertex. From the form of the
supercurrent (given in~\eqref{Sue}) we see that each $e^{\phi}$ factor is
accompanied by a $\partial X$, which in the amplitudes we are
interested in becomes a factor of the closed string momentum $k$. In
configuration space each factor of $k$ becomes a factor of $1/r$, $r$
being the radial coordinate in the $\mathbb{R}^4$. Thus amplitudes
with $m$ pairs of $V_\mu$, $V_{\bar\mu}$ vertices contribute to the geometric
backreaction with terms which decay at least as $1/r^m$ at large
distances.

We also have the standard open string loop expansion weighted by the
number of borders inserted in the diagrams and we need to
justify why we focus on disk amplitudes. The reason we can perform a perturbation expansion in the regime of parameters of interest is that the open string loop expansion parameter for the calculation of the value of the backreacted field at a radial distance $r$ (for a generic D$p$-brane) is 
\be \label{eq:epsilon}
\e ~=~ g_s N \left( \frac{\ap}{r^2} \right)^{\!\frac{7-p}{2}} 
\ee
where here $N$ counts the number of D1 or D5 branes. For fixed large
$g_s N$, $\e$ can be made arbitrarily small by choosing to examine the
fields at large enough $r$.

One can see that the quantity $\e$ controls the open string perturbation expansion as follows. Adding an extra border to the string worldsheet gives a factor of $g_s N$ since there are $N$ choices of which D-brane the open string endpoints can end on. It also introduces a loop momentum integral, two extra propagators, and reduces the background superghost charge by two units, requiring us to increase the picture of the vertex operators into a picture two units higher.

Qualitatively, each of these contributes as follows: At large distances, the loop momentum integral is dominated by the closed string channel, effectively resulting in an integral over the Dirichlet directions,
$\int d^{9-p} k$.
The two propagators bring two factors of $1/k^2$, and the picture-changing procedure brings a factor of $k^2$ as we have described. Thus all together we have an additional integral of the form
\be
\int d^{9-p} k \frac{1}{k^2}  ~ \sim ~   \frac{1}{r^{7-p}}\, 
\ee
and so restoring units of $\ap$ we indeed find that $\e$ is the appropriate dimensionless expansion parameter.

In the following sections we focus on the contributions which depend on all three charges of the microstate being present, and so vanish when any one charge is turned off; we refer to these as the `new' fields.
The new fields are not the only contributions appearing up to order $1/r^4$;
there are also contributions from diagrams with more borders which contribute at the same order.
However as mentioned in the Introduction, amplitudes with many disconnected borders are both more
difficult to derive and also less interesting. At large distances,
they should simply reproduce the contributions due to the non-linear
nature of gravity; these contributions are more easily calculated in
the low-energy limit, by solving the supergravity Killing spinor
equations (and equations of motion). The reason is that the momentum
of each closed string exchanged between a probe placed at large
distances and the D-brane bound state is very small. So the
contribution to the string amplitudes is dominated by world-sheets
which look like tree-level gravity Feynman diagrams where each
boundary represents a source. 
This is a diagrammatic representation for
the perturbative solution of the supergravity equations of motion
as performed in Section~\ref{sec:ansatz}.

Thus in the following we focus only on the leading contributions
at large distances which are induced by the amplitudes with one border
(see Figure~\ref{Dpfig}) and those with one border and one pair of
open vertices $V_\mu$, $V_{\bar\mu}$ (see Figure~\ref{Dmixfig}). This should be
sufficient to capture, in the supergravity solution corresponding to
any microstate, the interesting new terms up to order $1/r^4$.

\subsection{Amplitudes with one type of boundary} \label{sec:one_bdy}

\begin{figure}[htp]
\begin{center}
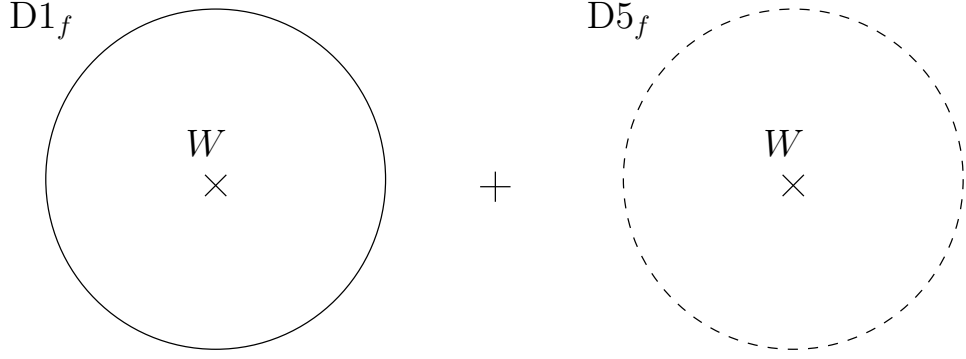
\end{center}
\caption{\label{Dpfig} The simplest non-trivial contributions to the
  one-point function of closed string state $W$: two disk diagrams where
  the border lies entirely on the D1 brane (for the first amplitude)
  or on the D5 brane (for the second amplitude).}
\end{figure}

The most direct way to derive the one-point functions from a disk with
only one type of boundary is to use the boundary state
formalism~\cite{DiVecchia:1997pr}. In our case we have two
contributions, depending on whether the boundary is ending on the D1
or D5 branes, see Figure~\ref{Dpfig}. These one-point couplings, and
the corresponding contributions to the background geometry, were
calculated in~\cite{Black:2010uq}, by using the boundary state for a
D-brane with a null wave derived
in~\cite{Hikida:2003bq,Blum:2003if,Bachas:2003sj}. Here we summarize
the derivation of the NS-NS fields in order to clarify a point on how
to separate the dilaton and the metric contribution which will be
useful in the following. For the R-R sector, we will just recall
the results of~\cite{Black:2010uq}.

Let us focus on the D1$_f$ diagram Figure~\ref{Dpfig}; the calculation
for the other contribution will be completely analogous. We can view
the wrapped D1-brane as as a collection of $n_w$ different D-brane
strands, with a non-trivial holonomy gluing these strands
together. Each strand carries a segment $f^i_{(s)}$, with
$s=1,\ldots,n_w$ of the full profile. The boundary state describing
the wrapped D1-brane can be expanded in terms of the closed string
perturbative states. The first terms of this expansions are
\begin{eqnarray} 
\label{eq:BdyStateStrands}
\ket{D1;f} & = &  -i \frac{ \k \, \t_{1} }{2} 
\sum_{s=1}^{n_w} \int \! du \int\limits_0^{2 \pi R} \! d{v} \int \!
\frac{d^4p_i}{(2 \pi)^4} \,e^{-i p_i f^i_{(s)}({v}) } 
\frac{c_0+\widetilde{c}_0}{2} 
\\ &  &\nonumber c_1 \widetilde{c}_1  \left[
  -\psi^\mu_{-\frac 12} ({}^{\rm t} \! R_{\rm D1})_{\mu\nu}
  \widetilde\psi^\nu_{-\frac 12} + \gamma_{-\frac 12} \widetilde\beta_{-\frac
    12} -  \beta_{-\frac 12} \widetilde\gamma_{-\frac 12} +\ldots
\right] \ket{u,v,p_i,0}_{-1,\widetilde{-1}}\, 
\end{eqnarray}
where $\t_{1}= [2\pi \ap g_s]^{-1}$ is the physical tension of a
D1-brane and where ${}^{\rm t}\!R_{\rm D1}$ is the transpose of the
reflection matrix $R_{\rm D1}$ given in \eq{eq:D1_bcs}. The ket
in~\eqref{eq:BdyStateStrands} represents a closed string state
obtained by acting on the $SL(2,C)$ invariant vacuum with a $e^{i p_i
  x^i}$ in the $\mathbb{R}^{4}$ directions. We also wrote the delta
functions on the $p_u$ and $p_v$ momenta as integrals in configuration
space $du,\,dv$. The boundary state enforces the
identification~\eqref{psii}, which in the
approximation~\eqref{eq:BdyStateStrands} holds just for the first
oscillator $\widetilde\psi^\mu_{-1/2}$.

The second line of~\eqref{eq:BdyStateStrands} contains all the
massless NS-NS states and we can separate the irreducible
contributions by taking the scalar product with each state, 
details of which are given in Appendix \ref{app:vertices}.
The dilaton state has the form (see~\eqref{stateop} for details):
\begin{equation}
  \label{stateop-1}
  \lim_{z\to 0} W_{\rm dil}^{(-2)} \ket{0} = \left(
\eta_{\mu\nu} c_1 \psi^\mu_{-\frac 12} \widetilde{c}_1 \widetilde{\psi}^\nu_{-\frac 12} +
c_1 \gamma_{-\frac 12} \widetilde{c}_1 \widetilde\beta_{-\frac 12} - 
c_1 \beta_{-\frac 12} \widetilde{c}_1 \widetilde\gamma_{-\frac 12}\right)
\ket{k}_{-1} \widetilde{\ket{k}}_{-1}\,
\end{equation}
while the graviton and the
B-field are given by the symmetric and the antisymmetric parts of
${\cal G}$ in ${\cal G}_{\mu\nu} c_1 \psi^\mu_{-1/2} \widetilde{c}_1
\widetilde{\psi}^\nu_{-1/2} \ket{k_i}_{-1,\widetilde{-1}}$ .

Some care is needed to read the diagonal components of the graviton,
as we have to select a state that is orthogonal
to~\eqref{stateop-1}. For these components we use
\begin{equation}
\nonumber %
\left[B  \left(
\eta_{\alpha\beta} c_1 \psi^\alpha_{-\frac 12} \widetilde{c}_1
\widetilde{\psi}^\beta_{-\frac 12} \right) + C \left(
c_1 \psi^i_{-\frac 12} \widetilde{c}_1 \widetilde{\psi}^i_{-\frac 12} +
c_1 \gamma_{-\frac 12} \widetilde{c}_1 \widetilde\beta_{-\frac 12} - 
c_1 \beta_{-\frac 12} \widetilde{c}_1 \widetilde\gamma_{-\frac 12}\right)
\right]\ket{k_i}_{-1,\widetilde{-1}} %\widetilde{\ket{k_i}}_{-1}
\end{equation}
where $B=-(7-p)/4$ and $C=(p+1)/4$ and we temporarily use the notation, for a generic D$p$-brane, that the indices $\alpha,\beta$ ($i$) are parallel (transverse) to the brane world-volume. The values of $B$ and $C$ ensure that this state has a vanishing scalar product with~\eqref{stateop-1}. We note that, if we analytically continue the
momentum $k_i$ so as to have $k^2=0$, the two round parenthesis are separately BRST-invariant.

We can now separate two types of contributions to the first diagram in
Figure~\ref{Dpfig}. The first type involves the diagonal part of the
reflection~\eq{eq:D1_bcs}, while the second one follows from the
elements in $R_{\rm D1}$ which depend on $f$. The result for the
diagonal terms is almost identical to the result for a static flat D-brane~\cite{DiVecchia:1997pr}: 
the one-point function for the
canonically normalized dilaton $\hat{\phi}$ is
\begin{equation}
%\label{eq:dilD1}
 \cA_{\rm dil}^{\rm D1}(k) ~=~ 
 - i \frac{\k \, \t_{1} }{2} V_u
\sqrt{2}\hat{\phi} \int\limits_0^{\,\,L_T}
 \! d\hat{v}  \,  e^{- i k \cdot f (\hat{v}) } ~,
\end{equation}
where $V_u$ is the (infinite) volume along the $u$ direction.
As we shall see, this term and the analogous D5-P contribution
are the only contributions to the dilaton for the D-brane configuration under
analysis. Similarly the one-point function for the diagonal components
$\hat{h}_{\mu\mu}$ of the canonically normalized graviton is
\begin{equation}
  \label{eq:graD1f0}
 - i \frac{\k \, \t_{1} }{2} V_u \int\limits_0^{\,\,L_T}
 \! d\hat{v}  \,  e^{- i k \cdot f  (\hat{v}) }  
 \left(- \frac{3}{2} %\frac{7-p}{4} 
 (-\hat{h}_{tt} +\hat{h}_{yy}) + %
 \frac{1}{2} %\frac{p+1}{4}
  (\hat{h}_{ii} +\hat{h}_{aa})\right)~.% 
\end{equation}
Here we already summed the contributions over all strands and so the
integrals over $v$ in each strand in~\eqref{eq:BdyStateStrands} have
been combined in a single integral over $\hat{v}$ extended from $0$ to
$L_T$.

The $f$-dependent terms in the reflection matrix switch on new couplings
with the off-diagonal terms of the metric\footnote{As we are
  considering a non-trivial profile only the $\mathbb{R}^4$
  directions, we do not have any contributions to the B-field.}. By
using the expression for the reflection~\eq{eq:D1_bcs}, one can see
that the complete graviton coupling induced by the first diagram in
Figure~\ref{Dpfig} is
\begin{eqnarray}
  \label{eq:graD1f}
\cA_{\rm gra}^{\rm D1}(k)& = &
 - i \frac{\k \, \t_{1} }{2} V_u \int\limits_0^{\,\,L_T}
 \! d\hat{v}  \,  e^{- i k \cdot f  (\hat{v}) }  
\left[-\frac{3}{2} (-\hat{h}_{tt} +\hat{h}_{yy}) \right.
\\ \nonumber & & \left. + \frac{1}{2}
   (\hat{h}_{ii} + \hat{h}_{aa}) -2 \hat{h}_{vv} |\dot{f}|^2 +
   4\hat{h}_{vi} \dot{f}^i\right] ~.
\end{eqnarray}
In the NS-NS sector, the expansion of the boundary state for a
D5-brane with a null wave $f$ is completely analogous to the D1
case~\eqref{eq:BdyStateStrands}, except for the appearance of the
D5-brane tension $\t_{5}= [(2\pi \sqrt{\ap})^5 \sqrt{\ap}g_s]^{-1}$
and the reflection matrix $R_{\rm D5}$ given in
~\eqref{eq:D5_bcs}. Thus we can read right away the contribution to
NS-NS couplings from the second diagram in Figure~\ref{Dpfig}:
\begin{eqnarray}
  \label{eq:dilD5}
 \cA_{\rm dil}^{\rm D5}(k) &=& 
 i \frac{\k \, \t_{5} }{2} V_u V_4
\sqrt{2}\hat{\phi} \int\limits_0^{\,\,L_T}
 \! d\hat{v}  \,  e^{- i k \cdot f  (\hat{v}) } ~,
\\  \label{eq:graD5f}
  \cA_{\rm gra}^{\rm D5}(k) & = & 
  - i \frac{\k \, \t_{5}}{2} V_u V_4 \int\limits_0^{\,\,L_T}
  \! d\hat{v}  \,  e^{- i k \cdot f  (\hat{v}) }  
  \left[-\frac{1}{2} (-\hat{h}_{tt} + \hat{h}_{yy}  + \hat{h}_{aa})  
  \right. \\ \nonumber & & \left. +\frac{3}{2} \hat{h}_{ii} -2 \hat{h}_{vv}
    |\dot{f}|^2 + 4\hat{h}_{vi} \dot{f}^i\right] 
\end{eqnarray}
where we recall the notation that $V_4$ is the volume of the compact space.

The details of the R-R calculation may be found
in~\cite{Black:2010uq}, here we just recall the results:
\begin{eqnarray}
  \label{eq:RRD1f}
  \cA_{\rm RR}^{\rm D1}(k) & = &
  - i \sqrt{2} \k \, \t_{1} 
  V_u \int\limits_0^{\,\,L_T}
  \! d\hat{v}  \,  e^{- i k \cdot f  (\hat{v}) }  
  \left[2 \hat{C}^{(2)}_{uv} + \hat{C}^{(2)}_{vi} \dot{f}^i\right]
\\   \label{eq:RRD5f}
  \cA_{\rm RR}^{\rm D5}(k) & = &
  - i \sqrt{2} \k \, \t_{5} 
   V_u V_4 \int\limits_0^{\,\,L_T}
  \! d\hat{v}  \,  e^{- i k \cdot f  (\hat{v}) }  
  \left[2 \hat{C}^{(6)}_{uv5678} + \hat{C}^{(6)}_{vi5678} \dot{f}^i\right] \,.
\end{eqnarray}

\subsection{Amplitudes with two types of boundary} \label{sec:2_boundaries}

\begin{figure}[htp]
\begin{center}
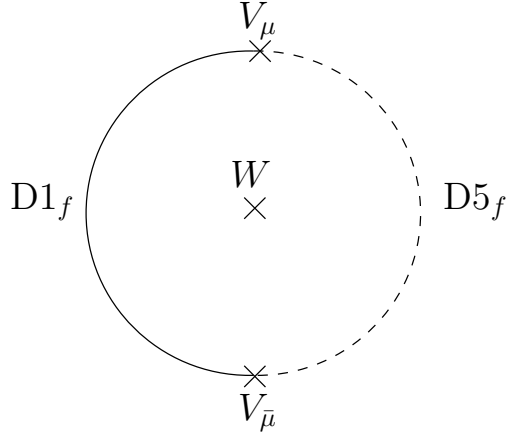
\end{center}
\caption{\label{Dmixfig} The simplest amplitude involving all three
  charges of the microstate: the topology of
  the worldsheet is that of a mixed disk diagram where part of the
  border lies on the D1 brane and part on the D5 brane.}
\end{figure}

We next calculate the contribution coming from diagrams with two types
of boundary. These diagrams were studied in the case of a D1-D5 bound
state without momentum charge in~\cite{Giusto:2009qq}; in this section
we introduce momentum charge by using the boundary conditions derived
earlier.  The world-sheet topology for these amplitudes is depicted in
Figure~\ref{Dmixfig} and involves a mixed disk with half its boundary
on a D1 and the other half on the D5 brane. Clearly this type of
diagram is absent for the naive D1/D5 superposition, where the fields
living on the D-brane world-volume are set to zero. On the contrary,
the configurations corresponding to D-brane bound states have a
non-zero vev for the massless fields in the spectrum of the open stings stretched between the D1 and
D5-branes. In our perturbative approach, these vevs are
described through the insertion of pairs of vertex operators, such as
$V_\mu$, corresponding to an open string stretching from the D1 to the
D5-branes, and $V_{\bar\mu}$, corresponding to an open string with the
opposite orientation. For the microstates we are interested in, these
vertex operators are given in~\eqref{Vop15}.

Thus the amplitude we need to calculate is
\begin{equation}\label{c}
  {\cal A}_{NS,R}^{\rm D1-D5} = \int \frac{\prod_{i=1}^4
    dz_i}{dV_{\mathrm{CKG}}} \, \left\langle V_\mu(z_1) \,
    W_{NS,R}^{(-k)}(z_2, z_3) \,
    V_{\bar \mu} (z_4) \,
  \right\rangle_f~,
\end{equation} 
where the subscript $f$ reminds that, in this disk correlator, the
identification between holomorphic and anti-holomorphic components
depends on the profile of the D-branes. In the parameterization where
the disk is mapped to the upper half of the complex plane, $z_1$
and $z_4$ are purely real as they represent the positions of the open
string vertices and lie on the boundary of the world-sheet,
while $z_2=\bar{z}_3$ is the position of the closed string vertex in
the interior of the world-sheet surface. In order to have a
non-trivial correlator we need to saturate the superghost charge of
the disk ($-2$): the two open string vertices together contribute half
of this total charge, so the closed string insertion has to carry
globally another $-1$ superghost charge.  Finally in~\eqref{c} we did
not keep track of the factors contributing to the overall
normalization. This normalization can be reabsorbed in the dictionary
between the string condensate~\eqref{mubarmu} and the supergravity
results and so it is not relevant for the comparison with the ansatz
of Section~\ref{sec:ansatz}.

\subsubsection{NS-NS Amplitude}

In the NS sector the holomorphic and the anti-holomorphic parts
of the vertices have integer superghost charge, and so the constraint on
the superghost charge forces us to work in an asymmetric picture; for
instance in the $(0,-1)$ picture, the closed string vertex operators
representing the emission of graviton or B-field is
\begin{equation}
W_{NS}^{(k)} =  {\cal G}_{\mu\nu}
 \left(\partial X^{\mu}_L -
\ii \;\! \frac{k}{2} \! \cdot \! \psi \, \psi^{\mu} \right)
 \,\ex{ \ii \frac{k}{2} \cdot X_L} (z)
\, \widetilde\psi^{\nu} 
\ex{-\widetilde\varphi} \ex{ \ii  \frac{k}{2} \cdot X_R }(\bar z) + \ldots \,,
\label{clNS0}
\end{equation}
where the dots stand for other terms that ensure the BRST invariance
of the vertex, but that do not play any role in the correlator under
analysis. For consistency, we should be able to choose the $(-1,0)$
picture, where the roles of the holomorphic and the anti-holomorphic
parts in~\eqref{clNS0} are swapped, or any linear combination of these
two choices. We show in Appendix~\ref{app:vertices} that all these
choices yield the same result.

The dilaton vertex we need to use is described in
Appendix~\ref{app:vertices} and is written in~\eqref{newdil-1}. This vertex has a
term involving the $(\xi,\eta)$ fields in the superghost
sector which does not contribute to the
correlator~\eqref{c}. The other term, which is relevant in our case,
has the same structure as the vertex~\eqref{clNS0}, but with 
${\cal G}_{\mu\nu}=\eta_{\mu\nu}$ and a symmetric linear combination of the
$(-1,0)$ and $(0,-1)$ structures. Since both structures yield the same
result in the correlator~\eqref{c}, we can use the simplified
form~\eqref{clNS0} for all massless NS-NS states, where the
polarization ${\cal G}$ is equal to the Minkowski metric for the
dilaton, while it is a symmetric (antisymmetric) tensor for the
graviton (B-field).

$SO(1,5)$ invariance, which is broken only by the boundary
conditions, dictates the form of the amplitude~\eqref{c} up to an
integral over the world-sheet punctures which was calculated
in~\cite{Giusto:2009qq}. By adapting that result to our case, we
obtain
\be
{\cal A}_{\rm NS}^{\rm D1 \mhyphen D5} 
 = - 2\sqrt{2} \pi   
  V_u \int\limits_0^{\,\,L_T}
 \! d\hat{v} e^{- i k \cdot f (\hat{v}) } 
k^K\, {\cal G}^{IJ} ({}^{\rm t}\!R)_J^{~M}\,v_{IMK} ~, \, \label{ampf0I-3} 
\ee
where we take ${}^{\rm t}\!R$ to be the transpose of the reflection matrix $R_{\rm D1}$ given in \eq{eq:D1_bcs}; 
we could also use $R = R_{\rm D5}$ and we would obtain the same result for the correlator under analysis.  The
$f$-dependent exponential factor follows from the zero-mode
part of the $e^{ikX_{L,R}/2}$ terms in the vertex
operator~\eqref{clNS0}, as a consequence of the Dirichlet boundary
conditions for the string coordinates $x^i=(X_L^i+X_R^i)/2$ in the
$\mathbb{R}^4$. The integral over $\hat{v}$ follows, as in the
previous section, from the zero-mode correlator along the Neumann
direction $v$, and, after combining the contributions from the
different D-brane strands, we can write the full amplitude as an
integral over the world-volume coordinate $\hat{v}$. 

Expanding the above amplitude~\eqref{ampf0I-3} for the D1-D5
condensate in which only the components $v_{uij}$ and $v_{vij}$ are
non-zero and $k$ is only in the $\mR^4$ directions, we obtain
\bea
{\cal A}_{\rm NS}^{\rm D1 \mhyphen D5} \! &\!\!\! = \!\!\! & 2\sqrt{2} \pi
 V_u \int\limits_0^{\,\,L_T}
 \! d\hat{v} e^{- i k \cdot f (\hat{v}) } k^l
\Big[
\left( \cG^{uj} + \cG^{ju} \right)  v_{ujl}  
+ \left( \cG^{vj} + \cG^{jv} \right) v_{vjl}
\label{eq:A_NS_expansion} \\ 
&& {} 
+ 4 \,\cG^{jv} \, v_{ujl} |\dot f(\hat{v})  |^2  
- 4 \,\cG^{ij} \, v_{uil} {\dot f}^j(\hat{v})   
- 2 \,\cG^{vv} \, v_{vil} {\dot f}^i(\hat{v})  
- 2 \,\cG^{uv} \, v_{uil} {\dot f}^i(\hat{v})  
\Big] \,,
\nonumber
\eea 
where the first line includes the two-charge D1-D5 contribution, and
the second one contains the new D1-D5-P contributions, which
vanish in each of the two-charge limits, {\em i.e.} if we set to zero
either the $v_{IJK}$ condensate or the profile $f^i$.

It is interesting to notice that the result~\eqref{eq:A_NS_expansion}
vanishes if we focus on the emission of a dilaton. As argued above, in
our amplitude one can effectively use $\eta_{\mu\nu}$ for the
dilaton polarization. There are two terms in the second line
of~\eqref{eq:A_NS_expansion} which can potentially contribute to the
dilaton amplitude, however, in our case, they cancel each other,
$$
 4 \,\eta^{ij} \, v_{uil} {\dot f}^j 
+ 2 \,\eta^{uv} \, v_{uil} {\dot f}^i  ~=~
 v_{uil} {\dot f}^i (4-4) ~=~ 0~,
$$
where we used $\eta^{ij}=\delta^{ij}$ and $\eta^{uv}=-2$. Thus the
only non-trivial contributions from the diagram in
Figure~\ref{Dmixfig} are for the graviton and B-field and can be read
from Eq.~\eqref{eq:A_NS_expansion} by using
\be
\cG_{\m\n} ~=~ \hat{h}_{\m\n} + \frac{1}{\sqrt{2}} \hat{b}_{\m\n} \,,
\ee
where, as before, $\hat{h}$ and $\hat{b}$ are the canonically
normalized supergravity fields. Thus we obtain the graviton coupling
\bea
{\cal A}_{\rm gra}^{\rm D1 \mhyphen D5}\! &=& - 2\sqrt{2} \pi 
 V_u \int\limits_0^{\,\,L_T}
 \! d\hat{v} e^{- i k \cdot f (\hat{v}) } k^l
\Big[
- 2 \hat{h}^{uj}  v_{ujl}  
- 2 \hat{h}^{vj}  v_{vjl}
\label{eq:D1D5gra} \\ 
&& {} 
- 4 \,\hat{h}^{jv} \, v_{ujl} |\dot f(\hat{v})  |^2  
+ 4 \,\hat{h}^{ij} \, v_{uil} {\dot f}_j(\hat{v})   
+ 2 \,\hat{h}^{vv} \, v_{vil} {\dot f}^i(\hat{v})  
+ 2 \,\hat{h}^{uv} \, v_{uil} {\dot f}^i(\hat{v})  
\Big]
\nonumber
\eea 
and the B-field coupling
\bea
{\cal A}_{\rm B}^{\rm D1 \mhyphen D5}\!  &=& 
 - 2 \pi V_u  \int\limits_0^{\,\,L_T}
 \! d\hat{v} e^{- i k \cdot f (\hat{v}) } k^l
\label{eq:D1D5b} \\ \! && {} \times
\Big[- 4 \,\hat{b}^{jv} \, v_{ujl} |\dot f(\hat{v})  |^2  
+ 4 \,\hat{b}^{ij} \, v_{uil} {\dot f}_j(\hat{v})   
+ 2 \,\hat{b}^{uv} \, v_{uil} {\dot f}^i(\hat{v})  
\Big]\,.
\nonumber
\eea 

\subsubsection{R-R Amplitude}

In the R-R sector the standard form for the massless closed string
vertices is, in the $(-1/2,-1/2)$ picture,
\begin{equation}
\label{clR0}
W_{R}^{(k)} = \frac{1}{8}\,  {\cal F}_{\hat{A}\hat{B}} 
\ex{-{\varphi\over 2}}\, S^{\hat{A}} \, \ex{ \ii \frac{k}{2} \cdot X_L} (z)
\,\ex{-\widetilde {\varphi\over 2}} \, \widetilde S^{\hat{B}}\, 
\ex{ \ii  \frac{k}{2} \cdot X_R}(\bar z) ~,
\end{equation}
where ${\cal F}_{\hat{A}\hat{B}}$ contains the fields strengths $F$ of
the R-R fields. It can be expanded on a basis of ten dimensional Gamma
matrices and contains a 1, a 3 and a self-dual 5-form
\be
{\cal F}_{\hat A \hat B} =\sum_{n=1,3,5}  
  \frac{1}{n!} F^{(n)}_{\mu_1..\mu_n}
  (C_{10}\Gamma_{(10)}^{\mu_1..\mu_n})_{\hat{A}\hat{B}}
  \,.  \label{f135}  
\ee
The standard relation between the field strength $F^{(n)}$ and its
$U(1)$ gauge potentials $C^{(n-1)}$ reads in momentum space as follows
\be \label{eq:F_to_C}
F^{(n)}_{I_1..I_n}=n \,\ii\, k_{[I_1} \,
\hat{C}^{(n-1)}_{I_2..I_n]}~. 
\ee
The amplitude we next calculate is again~\eqref{c}, now with the R-R vertex~\eqref{clR0} inserted. The holomorphic and
anti-holomorphic spin fields are identified via the spinor
representation of the reflection matrix~\eqref{eq:cR_D1P}; after this
identification, one obtains the same fermionic correlator, with four
spin fields, of the two-charge case~\cite{Giusto:2009qq} and so the modified reflection matrix brings all the new terms with respect to the two-charge calculation. 

Since the open string condensate $\bar \mu^{(A} \mu^{B)}$ under
consideration is invariant under the $SO(4)$ Lorentz group of the
$T^4$ torus, we can restrict ourselves to $SO(4)$ invariant components
of ${\cal F}_{\hat A \hat B}$, which, by using the conventions of
Appendix~\ref{app:conventions}, are ${\cal F}_{AB[\dot\alpha
  \dot\beta]}$ , ${\cal F}^{AB[\alpha \beta]}$. In addition the RR
components ${\cal F}^{AB[\alpha \beta]}$ can be discarded by noticing
that the only $SO(6)$ singlet $\epsilon_{ABCD} \mu^{A} \bar\mu^{B}\,
{\cal F}^{CD [\alpha \beta]} $ vanishes for the symmetric open string
condensate $\bar \mu^{(A} \mu^{B)}$ we consider.
As a result, the open string condensate under analysis contributes only
to the emission of
\beq
\label{clR}
W_{R}^{\rm ef} = \frac{1}{8}\,({\cal F}\cR)_{A B} \epsilon_{\dot\alpha\dot\beta}
 \ex{-{\varphi\over 2}}\, S^{A} \, S^{\dot\alpha} \,  (z)
 \,\ex{- {\varphi\over 2}} \, S^{B}\, S^{\dot \beta}(\bar z) ~,
\eeq
with $\cR = \cR_{\rm D1}$ (or alternatively $\cR = \cR_{\rm D5}$) and where
\bea
{\cal F}_{AB} &=&   \frac{1}{2\,  4!}  F^{(5)}_{Iabcd}( C_{10}
\Gamma^{Iabcd}_{(10)})_{AB}{}^{\dot\alpha}_{\dot\alpha}+ 
\sum_{n=1,3,5} \frac{1}{2\,  n!}  
F^{(n)}_{I_1..I_n} (C_{10} \,
\Gamma^{I_1..I_n}_{(10)})_{AB}{}^{\dot\alpha}_{\dot\alpha}\nn\\ 
&=&  F^{(5)}_{I 5678}
( C \,\Gamma^{I})_{AB} +\sum_{n=1,3,5} \frac{1}{  n!}   
F^{(n)}_{I_1..I_n} (C \,\Gamma^{I_1..I_n})_{AB} \,. \label{rrforms}
\eea
In the last line above we have used the facts that $\Gamma^{5678}_{(10)} =
-\Gamma^{3\bar 3 4 \bar 4}_{(10)}=1_{(6)}\otimes (-\gamma^{ND})$ and
that $-\gamma^{ND}$ is $1$ on the indices $\dot{\alpha}$ (see \eqref{gGamma}).
Lorentz invariance again fixes the form of ${\cal A}_R$ up to a
constant calculated in~\cite{Giusto:2009qq}, giving
\be
{\cal A}_R^{\rm D1\mhyphen D5} ~=~ \frac{i \pi}{2} V_u 
 \int\limits_0^{\,\,L_T} \! d\hat{v} \, e^{- i k \cdot f (\hat{v}) } 
\bar \mu^A(C^{-1} {\cal F} \cR_0 C^{-1})_{AB}\, \mu^B~,
\ee
where $\cR_0$ is the $SO(1,5)$ part of the spinorial reflection matrix
$\cR$ given in~\eq{eq:cR_D1P}, i.e.
\be \label{eq:cR_D1P_SO15}
\cR_0 ~=~ \frac{1}{2} \G^{uv} + \dot{f}^i(v)\G^{iv} \,.
\ee
The couplings coming from the $\ha \G^{uv}$ term are the two-charge
D1-D5 contributions; these were calculated in~\cite{Giusto:2009qq} and
contribute first at order $1/r^3$. The terms coming from the
$\dot{f}^i(v)\G^{iv}$ term are new D1-D5-P terms and contribute first
at order $1/r^4$.

Expanding the amplitude in terms of the R-R gauge potentials
using \eq{rrforms} and \eq{eq:F_to_C} we find the following coupling
to the canonically normalized R-R potentials:
\bea 
{\cal A}_R^{\rm D1\mhyphen D5} \!\!& = \!\!& 4 \pi V_u 
 \int\limits_0^{\,\,L_T} \! d\hat{v} \, e^{- i k \cdot f (\hat{v}) } k^l
\Bigg[ ({\hat C}^{(2)})^{vj} \, v_{vlj} - ({\hat C}^{(2)})^{uj} \,
v_{ulj}  \cr 
&&\!\!\!\! {} - 2 {\hat C}^{(0)} \, v_{ulj} {\dot f}^j(\hat{v})  + ({\hat C}^{(2)})^{uv} \, v_{ulj} {\dot f}^j(\hat{v})  
-2 ({\hat C}^{(2)})^{ij} \left(  v_{uli}\,{\dot f}^j(\hat{v}) -  v_{ulj}\,{\dot f}^i(\hat{v}) \right)  \cr
&&\!\!\!\! {} - 2 ({\hat C}^{(4)})^{5678} \, v_{ulj} {\dot f}^j(\hat{v}) 
+ ({\hat C}^{(4)})^{uvij} \left( v_{uli}\,{\dot f}^j(\hat{v}) -  v_{ulj}\,{\dot f}^i(\hat{v}) \right) 
\Bigg] \qquad\qquad\qquad \label{eq:A_R_expansion}
\eea
where the first line is the two-charge D1-D5 contribution, and
the other terms are the new D1-D5-P contributions, which
vanish in each of the two-charge limits.

\subsection{Geometry from string amplitudes}

We now use the amplitudes calculated in the previous section to read
off the couplings with the canonically normalized fields and then to
derive the geometric backreaction of the microstate at large
distances. As usual the couplings between the D-brane configuration
and the perturbative states is given simply by the first variation of
the string amplitude
\begin{align}  \label{eq:diff_fields} 
\hat{h}_{\m\n} & = \frac{1}{2}  
\frac{\delta {\cal A}_{\rm NS}}{\delta \hat{h}^{\m\n} } 
~~&(\m < \n) \,,&  &
\hat{h}_{\m\m} &= \frac{\delta {\cal A}_{\rm NS}}{\delta \hat{h}^{\m\m}}  
& & (\mathrm{no~sum~over}  ~\m) ~, 
\\ \label{eq:diff_fields2} 
\hat{b}_{\m\n} &= \frac{\delta {\cal A}_{\rm NS}}{\delta
  \hat{b}^{\m\n}} ~~ & (\m < \n) \,, & &
\hat{C}^{(n)}_{\mu_1\ldots\mu_n} & = \frac{\delta {\cal A}_{\rm R}}{
  \delta \hat{C}^{(n) \mu_1\ldots\mu_n}}
& & (\mu_1<\mu_2\ldots<\mu_n)\,,
\end{align}
where all the fields are set to zero after the variations. The NS-NS
amplitude is the combination of the results 
in~\eqref{eq:graD1f},~\eqref{eq:graD5f} and~\eqref{eq:A_NS_expansion},
plus higher order corrections
\begin{equation}
  \label{eq:ans}
  {\cal A}_{\rm NS} = \frac{1}{V_4 V_u 2\pi R} \left(
{\cal A}_{\rm NS}^{\rm D1} + {\cal A}_{\rm
    NS}^{\rm D5} + {\cal A}_{\rm NS}^{\rm D1-D5} + \ldots \right)~, 
\end{equation}
where the prefactor is needed to cancel the volume of the directions
where the momentum of the emitted closed string is set to zero.  A
similar equation holds in the R-R sector where one must
combine~\eqref{eq:RRD1f},~\eqref{eq:RRD5f}
and~\eqref{eq:A_R_expansion}.

If we indicate with $a_{\mu_1\ldots\mu_n}(k)$ a generic coupling
appearing in~\eqref{eq:diff_fields} and~\eqref{eq:diff_fields2}, then the we can
derive the geometric backreaction of the configuration by sewing a
standard propagator $-i/k^2$ and then taking the Fourier transform to rewrite
the result in configuration space,
\beq\label{fourier}
a_{\mu_1\ldots\mu_n}(x) = \int \frac{d^4 k}{(2\pi)^4}
\left(-\frac{\ii}{k^2}\right)  
a_{\mu_1\ldots\mu_n}(k)\,
\ex{\ii k \cdot x} ~.
\eeq
A common feature of the string couplings derived in the previous
section is the presence of a $f$-dependent exponential factor ($e^{-i k \cdot f}$), which combines with the $e^{i k \cdot x}$ Fourier
transformation. Thus the contributions of Section~\ref{sec:one_bdy}
and of~\ref{sec:2_boundaries} will involve the following integrals~\eqref{eq:FT0}
and~\eqref{eq:FT} respectively,
\bea \label{eq:FT0}
\int \frac{d^4 k}{(2\pi)^4} \left(- \frac{i}{k^2} \right) e^{i k^i (x^i - f^i) } &=& 
{ -i\over 4\pi^2}\frac{1}{|x^i - f^i|^2 }~,
\\
\label{eq:FT}
\int \frac{d^4 k}{(2\pi)^4} \left(- \frac{i}{k^2} \right) k^l e^{i k^i (x^i - f^i) } &=& 
{ 1\over 2\pi^2}\frac{x^l - f^l}{|x^i - f^i|^4 }~
\eea
where in the above we have used the abuse of notation 
\be \label{eq:abuse}
|x^i-f^i|^2 = \sum\limits_{i=1}^4 (x^i-f^i)	^2 \,.
\ee
We now wish to compare the string result with the ansatz of
Section~\ref{sec:ansatz}, so we must change from the canonically
normalized (hatted) fields which have propagators $1/k^2$ to the
fields appearing in the supergravity action
\be \label{eq:normalization}
g~=~\eta+2 \kappa \hat{h} ~, \quad B ~=~ \sqrt{2}\kappa \hat{b}~,
\quad \phi ~=~ \sqrt{2}\kappa \hat{\phi}~, 
\quad C^{(n)} ~=~ \sqrt{2} \kappa \hat{C}^{(n)} ~.
\ee
We then use Eqs.~\eqref{eq:graD1f}, \eqref{eq:graD5f}
and~\eqref{eq:D1D5gra} in Eq.~\eqref{fourier} to derive the metric
induced by our D-brane configuration. We thus obtain the following
dilaton and metric components (in the Einstein frame):
\bea\label{stringphi}
\ex{\phi} &=& 1 + \frac{1}{L_T} \int\limits_0^{\,\,L_T}
\frac{(Q_1- Q_5)}{2 |x^i - f^i|^2}d\hat{v} ~,
\\
g_{uj} &=&  
\frac{1}{L_T} \, \int\limits_0^{\,\,L_T} \! \mathbf{v}_{ulj} 
\frac{x^l - f^l}{|x^i - f^i|^4 } d\hat{v} \,,\\
g_{vj} &=& \frac{1}{L_T}
\int\limits_0^{\,\,L_T} \left[ 
\frac{x^l - f^l}{|x^i - f^i|^4 }
\left( \mathbf{v}_{vlj} + 2 |\dot f^i|^2 \mathbf{v}_{ulj} \right) 
- \frac{(Q_1+ Q_5) \dot{f_j}}{|x^i - f^i|^2} \right]d\hat{v} \,,\\
g_{ij} &=&  \delta_{ij} \left(1 + \frac{1}{L_T} \int\limits_0^{\,\,L_T}
\frac{(Q_1+ 3 Q_5)}{4 |x^i - f^i|^2}d\hat{v}  \right)
\label{eq:g_ij_r4-5} \\ \nonumber  &-& \frac{2}{L_T} \,
 \int\limits_0^{\,\,L_T} \left( \frac{{\dot f}_{j} (x^l -
   f^l)}{|x^i - f^i|^4 } \mathbf{v}_{uli}  + 
 \frac{{\dot f}_{i} (x^l -    f^l)}{|x^i - f^i|^4 } 
 \mathbf{v}_{ulj} \,\right) d\hat{v}
\\
g_{ab} &=& \delta_{ab} \left(1 + \frac{1}{L_T} \int\limits_0^{\,\,L_T} 
\frac{(Q_1 - Q_5)}{4 |x^i - f^i|^2}d\hat{v} \right)
\\
g_{uv} &=&  - \frac{1}{2} +  \frac{1}{L_T} \int\limits_0^{\,\,L_T}
\frac{(3 Q_1+ Q_5)}{8 |x^i - f^i|^2}d\hat{v}
- \frac{1}{L_T}
\int\limits_0^{\,\,L_T} \!  \mathbf{v}_{ulj} \frac{{\dot f}^{j} (x^l -
   f^l)}{|x^i - f^i|^4 } d\hat{v}  \,, \\ 
g_{vv} &=&  \frac{1}{L_T}
\int\limits_0^{\,\,L_T} \left[  -2 
\mathbf{v}_{vlj} \frac{{\dot f}^{j} (x^l - f^l)}{|x^i - f^i|^4 } 
+ \frac{(Q_1+ Q_5) |\dot{f^j}|^2}{|x^i - f^i|^2}
\right]\,  d\hat{v}~\label{stringgvv}
\eea
where in order to make the equations more readable, we have absorbed some factors
in the $\bar\mu \mu$ condensate 
\begin{equation}
  \mathbf{v}_{IJK} ~=~ -\frac{2 \sqrt{2} n_w \kappa}{\pi V_4} v_{IJK} 
\end{equation}
and we have introduced the standard combinations $Q_1$ and $Q_5$,
\begin{equation} \label{Q1Q5}
Q_1 =  \frac{n_w \tau_1 \kappa^2}{2\pi^2 V_4} = \frac{(2\pi)^4 g_s \ap^3
  n_w}{V_4}~, ~~~~~
Q_5 = \frac{n_w \tau_5 \kappa^2}{2\pi^2} = \ap g_s n_w ~.
\end{equation}
In the above, if we set to zero the condensate $v_{IJK}$ for the open
string stretched between the D1 and the D5 branes we recover the a
linear combination of two-charge solutions D1/P and D5/P, as already
verified in~\cite{Black:2010uq}. Another limit consists of switching
off the momentum charge by setting $f=0$. In this case we recover the
solutions appropriate for the two-charge microstates D1/D5. Of
particular interest are the new contributions that vanish in both
limits.

In order to obtain the large distance behaviour of the supergravity
fields listed above, it is sufficient to expand the denominators for
$x^i\gg f^i$. Some of the leading contributions vanish because of the
periodicity of the profile, which implies $\int_0^{L_T} f^i(\hat{v})
d\hat{v}= 0$ meaning that the first non-trivial corrections are proportional to
the moment of the wave $\int_0^{L_T} \dot{f}^i f^j d\hat{v}\,$. As
discussed in some details at the beginning of this section, the
$1/r^4$ behaviour obtained in this way should be universal for all
microstates which have the same momentum profile on the D1 and the
D5-branes. Here we decided to keep the exact dependence on $f^i$, as
it follows from the string computation, since these formulae are
relevant for a smaller class of microstate where the condensate
$v_{IJK}$ is small (or in other words, states which are localized near
the origin of the classical Higgs branch). 
This information will
provide a useful guide when generalizing the perturbative $1/r$
expansion to a full non-linear supergravity solution.
For this reason,
in the following, we will keep the exact dependence of the string
results on the momentum profile.

In a similar way, from Eq.~\eqref{eq:D1D5b} we obtain the non-zero
components of the B-field:
\bea
B_{vj} &=& \frac{2}{L_T} 
\int\limits_0^{\,\,L_T} |\dot{f}|^2\,\mathbf{v}_{ujl} \, 
\frac{x^l - f^l}{|x^i - f^i|^4 } d\hat{v}
\label{bvj}
\,,\qquad\qquad~\\
B_{ij} &=& -\frac{2}{L_T} 
\int\limits_0^{\,\,L_T}
\left(\mathbf{v}_{uli} \, \frac{{\dot f}_{j} (x^l - f^l)}{|x^i - f^i|^4} 
- \mathbf{v}_{ulj} \,  \frac{{\dot f}_{i} (x^l - f^l)}{|x^i - f^i|^4}\right)
d\hat{v} 
\label{bij}
\,,  \\ 
B_{uv} &=&  -\frac{1}{L_T} 
\int\limits_0^{\,\,L_T}
\, \mathbf{v}_{ulj}  \frac{{\dot f}^{j} (x^l - f^l)}{|x^i - f^i|^4 } 
d\hat{v} \,.
\label{buv}
\eea
Notice that all these contributions are new in the sense that they disappear in the two-charge limits where either $f^i$ or
$v_{IJK}$ are set to zero. On the contrary we have seen that the dilaton
does not receive such new contributions from the mixed disk diagram in
Figure~\ref{Dpfig}.

Following the same approach with the R-R fields, we use
Eqs.~\eqref{eq:RRD1f},~\eqref{eq:RRD5f} and~\eqref{eq:A_R_expansion}
in Eq.~\eqref{fourier} to derive the backreaction of the microstate
under analysis in this sector. We find the nonzero R-R fields:
\bea
C^{(0)} &=&    \frac{2}{L_T} \,\int\limits_0^{\,\,L_T} \!  \mathbf{v}_{ulj}
\frac{{\dot f}^{j} (x^l - f^l)}{|x^i - f^i|^4 } d\hat{v}  \,, \\ 
\label{stringC2vj}
C^{(2)}_{vj} &=& {}  \frac{1}{L_T} \, \int\limits_0^{\,\,L_T} 
\left[- \mathbf{v}_{vlj} \frac{x^l - f^l}{|x^i - f^i|^4 } 
+ Q_1 \frac{\dot{f_j}}{|x^i - f^i|^2}
\right] d\hat{v} \,, \qquad \quad \\ 
C^{(2)}_{uj} &=&  \frac{1}{L_T} \, 
\int\limits_0^{\,\,L_T} \mathbf{v}_{ulj} \frac{x^l - f^l}{|x^i -
  f^i|^4 } d\hat{v} \,, \\  
C^{(2)}_{uv} &=&   \frac{1}{L_T}
\,\int\limits_0^{\,\,L_T} \left[-\mathbf{v}_{ulj} \frac{{\dot f}^{j}
    (x^l -  f^l)}{|x^i - f^i|^4}  + \frac{Q_1}{2 |x^i - f^i|^2}\right]
    d\hat{v}  \,, \\   
\label{stringC2ij}
C^{(2)}_{ij} &=&  \frac{2}{L_T} \int\limits_0^{\,\,L_T} \left[
  \mathbf{v}_{uli} \, \frac{{\dot f}_{j} (x^l - f^l)}{|x^i - f^i|^4} -
  \mathbf{v}_{ulj} \, \frac{{\dot f}_{i} (x^l - f^l)}{|x^i -
    f^i|^4}\right] d\hat{v} \,,  \\  
C^{(4)}_{uvij} &=& {} - \frac{1}{L_T}
\int\limits_0^{\,\,L_T} \left[ \mathbf{v}_{uli} \, \frac{{\dot f}_{j} (x^l -
    f^l)}{|x^i - f^i|^4} - \mathbf{v}_{ulj} \,  \frac{{\dot f}_{i} (x^l -
    f^l)}{|x^i - f^i|^4}\right] d\hat{v} \,,  \\  
C^{(4)}_{5678} &=& \frac{2}{L_T} \,\int\limits_0^{\,\,L_T} \!
\mathbf{v}_{ulj} \frac{{\dot f}^{j} (x^l - f^l)}{|x^i - f^i|^4 }
d\hat{v}  \,, \\  
C^{(6)}_{vj5678} &=&  \frac{Q_5}{L_T} \,
\int\limits_0^{\,\,L_T}  \frac{\dot{f_j}}{|x^i - f^i|^2} d\hat{v}
\,, \\  
C^{(6)}_{uv5678} &=&  \frac{1}{L_T} \,
\int\limits_0^{\,\,L_T}  \frac{Q_5}{2 |x^i - f^i|^2} d\hat{v}
\,. 
\eea
In the next section we read off from these fields the string
contribution to the fields parameterizing the supergravity ansatz of
Section~\ref{sec:ansatz}.

\section{Comparison to supergravity} \label{sec:comparison}

We will now verify that the fields derived from the string amplitudes
satisfy the supergravity constraints obtained in Section
\ref{sec:ansatz}.

The supergravity analysis was performed in the large $r$ limit,
keeping only terms up to order $1/r^4$. One should thus apply the
supergravity equations to the large $r$ expansion of the string results
of the previous section.
It turns out, however, that
one can keep the full $r$ dependence of the string results and still
satisfy\footnote{This happens because the approximate
  constraints of Section \ref{sec:ansatz} are valid up to terms of
  second order in the condensate $\mathbf{v}_{IJK}$ and thus in the
  same approximation in which the string results have been derived.} 
the approximate supergravity equations of Section \ref{sec:ansatz}.
This is what we will show in the following. We remind the reader that 
full $r$ dependence of the supergravity fields is meaningful in describing
the small $g_s N$ and small $\mathbf{v}_{IJK}$ limit, i.e. the weak gravity regime and 
the region of the Higgs branch 
infinitesimally close to its intersection with the Coulomb branch. If one 
is interested in the full black hole regime (large $g_s N$ and finite $\mathbf{v}_{IJK}$), 
one should keep only the large $r$ limit (up to $1/r^4$ order) of the results we present below.

In order to make equations more compact, it is useful to define the
the following integrals 
\be
\label{integraldef}
\mathcal{I} = \frac{1}{L_T} \int_0^{L_T} \!\!d\hat v\,\frac{1}{|x^i-f^i|^2}\,,\,\,\, \widetilde{\mathcal{I}} = \frac{1}{L_T} \int_0^{L_T} \!\!d\hat v\,\frac{|\dot{f}^j|^2}{|x^i-f^i|^2}\,,\,\,\, \mathcal{I}_j = \frac{1}{L_T} \int_0^{L_T} \!\!d\hat v\,\frac{\dot{f}_j}{|x^i-f^i|^2}\,.
\ee
The properties
\be\label{integralprop}
\partial_i^2\, \mathcal{I}= \partial_i^2\, \widetilde{\mathcal{I}}=\partial_i^2\, \mathcal{I}_j=0\,,\quad \partial_i\, \mathcal{I}_i=0 
\ee
easily follow from the definitions  above and the fact that $f_i(\hat v)$ is a periodic function.

We first extract from the string results of Section \ref{sec:backreaction} the metric functions used to parameterize the general supergravity ansatz of Section \ref{sec:ansatz} and then verify that they obey the constraints from supersymmetry and the equations of motion. In doing this we will only keep terms up to first order in the condensate $\mathbf{v}_{IJK}$ and in the $Q_1$ and $Q_5$ charges. 

As the supergravity ansatz is given in the string frame, it is useful to translate the string results for the metric, given in Einstein frame, into string frame. At our order of approximation, if we denote by $\eta_{\mu\nu} + h_{\mu\nu}$ and by  $g_{\mu\nu}$ the string frame and Einstein frame metrics, one has
\be
\eta_{\mu\nu} + h_{\mu\nu}= g_{\mu\nu}+\frac{1}{2}\,\eta_{\mu\nu}\,\phi\,.
\ee
From eqs.~(\ref{stringphi})-(\ref{stringgvv}), one then finds that the world-sheet prediction for the metric in string frame is
\bea
h_{uj}&=& -\frac{1}{2}\,\mathbf{v}_{ulj}\,\partial_l \mathcal{I}\,,\\
h_{vj}&=& -\frac{1}{2} \,\mathbf{v}_{vlj}\,\partial_l  \mathcal{I}- \mathbf{v}_{ulj}\,\partial_l \widetilde{\mathcal{I}}- (Q_1+Q_5)\,\mathcal{I}_j\,,\\
h_{ij}&=&\frac{1}{2}\,(Q_1+Q_5)\,\mathcal{I}\,\delta_{ij} + \mathbf{v}_{uli}\,\partial_l \mathcal{I}_j +  \mathbf{v}_{ulj}\,\partial_l \mathcal{I}_i\,,\\
h_{ab}&=&\frac{1}{2}\, (Q_1 - Q_5)\,\mathcal{I}\,\delta_{ab}\,,\\
h_{uv}&=& \frac{1}{4}\,(Q_1+Q_5)\,\mathcal{I} +\frac{1}{2}\,\mathbf{v}_{ulk}\,\partial_l \mathcal{I}_k \,,\\
h_{vv}&=& \mathbf{v}_{vlk}\,\partial_l \mathcal{I}_k+(Q_1+Q_5)\,\widetilde{\mathcal{I}}\,.
\eea
We should also dualize the RR 6-form computed on the string side into a 2-form:
\bea
&&\!\!\!\!\!\!\!\!\!\!\!C^{(6)} = Q_5 \,\Bigl(\frac{1}{2}\,\mathcal{I}\,du\wedge dv + \mathcal{I}_i \, dv\wedge dx_i\Bigr) \,\wedge dz^4 \quad \Rightarrow\nonumber\\
&&\quad d\widehat{C}^{(2)} = -* d C^{(6)} = Q_5\,\epsilon_{ijkl}\, (\partial_l \mathcal{I}\,dx^k +\partial_k\mathcal{I}_l \, dv) \wedge dx^i\wedge dx^j\,.
\eea
One can check that, thanks to (\ref{integralprop}), the forms 
\be
\epsilon_{ijkl}\,\partial_l\mathcal{I} \, dx^i\wedge dx^j\wedge dx^k\quad  \mathrm{and}\quad  \epsilon_{ijkl}\,\partial_k\mathcal{I}_l \, dx^i\wedge dx^j
\ee
are $d$-closed, and hence one can define a 2-form $\frac{1}{2}\,\widehat{\mathcal{I}}_{ij}\,dx^i\wedge dx^j$ and a 1-form  $\widehat{\mathcal{I}}_i \,dx^i$ such that
\be\label{Ihat}
d\Bigl(\frac{1}{2}\, \widehat{\mathcal{I}}_{ij} \,dx^i\wedge dx^j\Bigr) = \epsilon_{ijkl}\,\partial_l\mathcal{I} \, dx^i\wedge dx^j\wedge dx^k\,,\quad \,d (\widehat{\mathcal{I}}_i \,dx^i )= \epsilon_{ijkl}\,\partial_k\mathcal{I}_l \, dx^i\wedge dx^j\,.
\ee
Then the dual of $C^{(6)}$ is 
\be
\widehat{C}^{(2)} = Q_5\,\Bigl(\frac{1}{2}\,\widehat{\mathcal{I}}_{ij} \,dx^i\wedge dx^j - \widehat{\mathcal{I}}_i \,dv\wedge \,dx^i\Bigr)\,.
\ee
This gives additional contributions to the $C^{(2)}_{vj}$ and $C^{(2)}_{ij}$of eqs. (\ref{stringC2vj}),(\ref{stringC2ij}) , so that in total one has
\bea
C^{(2)}_{vj} &=& \frac{1}{2}\, \mathbf{v}_{vlj}\,\partial_l \mathcal{I} + Q_1\,\mathcal{I}_j - Q_5 \,\widehat{\mathcal{I}}_j\,,\\
C^{(2)}_{ij} &=& - \mathbf{v}_{uli}\,\partial_l \mathcal{I}_{j}-  \mathbf{v}_{ulj}\,\partial_l \mathcal{I}_{i}+Q_5\,\widehat{\mathcal{I}}_{ij}\,.
\eea

We can now compute the various metric coefficients that appear in the supergravity ansatz (\ref{NSNSsusy}), (\ref{RRsusy}), (\ref{f5susy}) :
\bea\label{functionstart}
Z_1 \!\!& =&\!\! 1+ 2\, h_{uv} + h_{aa} = 1+ Q_1 \,\mathcal{I} + \mathbf{v}_{ulk}\, \partial_l \mathcal{I}_k\,,\\
Z_2 \!\!&=&\!\! 1+2\, h_{uv} - h_{aa}=1+ Q_5 \,\mathcal{I} + \mathbf{v}_{ulk}\, \partial_l \mathcal{I}_k\,,\\
Z_3 \!\!&=&\!\! 1 + h_{vv}= 1+(Q_1+Q_5)\,\widetilde{\mathcal{I}}+\mathbf{v}_{vlk}\,\partial_l \mathcal{I}_k \,,\\
a_3\!\!&=&\!\! -2\, h_{ui}\,dx^i = \mathbf{v}_{uli}\,\partial_l \mathcal{I}\,dx^i\,,\\
k\!\!&=&\!\! - (h_{ui} + h_{vi})\,dx^i =\!\Bigl[(Q_1 + Q_5)\,\mathcal{I}_i + \frac{1}{2} (\mathbf{v}_{uli}+\mathbf{v}_{vli}) \,\partial_l \mathcal{I} + \mathbf{v}_{uli}\,\partial_l \widetilde{\mathcal{I}} \Bigr] dx^i ,\quad\phantom{}\\
ds^2_4\!\!&=&\!\! (\delta_{ij} + h_{ij} -2\, h_{vu} \,\delta_{ij})\,dx^i dx^j \nonumber\\
&=&\!\! [\delta_{ij} + \mathbf{v}_{uli}\,\partial_l \mathcal{I}_j +\mathbf{v}_{ulj}\,\partial_l \mathcal{I}_i - \delta_{ij} \,\mathbf{v}_{ulk}\,\partial_l \mathcal{I}_k]\,dx^i dx^j\,,\label{basemetric}\\
D \!\!&=&\!\! e^{2\phi} = 1+ (Q_1-Q_5)\,\mathcal{I}\,,\\
b_0\!\!&=&\!\!-2\,B_{uv} = - \mathbf{v}_{ulk}\,\partial_l \mathcal{I}_k\,,\\
b_1\!\!&=&\!\! (B_{ui}-B_{vi})\,dx^i = \mathbf{v}_{uil}\,\partial_l \widetilde{\mathcal{I}}\,dx^i\,,\\
\widetilde b_1 \!\!&=&\!\! -(B_{ui}+B_{vi})\,dx^i=\mathbf{v}_{uil}\,\partial_l \widetilde{\mathcal{I}}\,dx^i\,,\\
b_2 \!\!&=&\!\! \frac{1}{2}\,B_{ij}\,dx^i \wedge dx^j = \mathbf{v}_{uli}\,\partial_l \mathcal{I}_j\,dx^i\wedge dx^j\,,\\
c&=& C^{(0)} = -\mathbf{v}_{ulk}\,\partial_l \mathcal{I}_k\,,\\
\widetilde Z_1 &=& 1 + 2 \,C^{(2)}_{uv} = 1 + Q_1\,\mathcal{I}+ \mathbf{v}_{ulk}\,\partial_l \mathcal{I}_k\,,\\
a_1 &=& (-h_{ui}-h_{vi}+C^{(2)}_{ui}-C^{(2)}_{vi})\,dx^i = Q_5\,(\mathcal{I}_i + \widehat{\mathcal{I}}_i)\,dx^i +\mathbf{v}_{uli}\,\partial_l \widetilde{\mathcal{I}}\,dx^i \,,\\
\widetilde a_1 &=& (h_{ui}-h_{vi}-C^{(2)}_{ui}-C^{(2)}_{vi})\,dx^i = Q_5\,(\mathcal{I}_i + \widehat{\mathcal{I}}_i)\,dx^i +\mathbf{v}_{uli}\,\partial_l \widetilde{\mathcal{I}}\,dx^i \,,\\
\widetilde\gamma_2 &=& \frac{1}{2}\,C^{(2)}_{ij}\,dx^i \wedge dx^j=  \frac{1}{2}\, Q_5\,\widehat{\mathcal{I}}_{ij}\,dx^i\wedge dx^j- \mathbf{v}_{uli}\,\partial_l \mathcal{I}_{j}\,dx^i\wedge dx^j \,,\\
f&=& C^{(4)}_{5678} = - \mathbf{v}_{ulk}\, \partial_l \mathcal{I}_k\,.\label{functionend}
\eea
Finally we need to derive the 0-forms $\widetilde{b}_0$ and $\widetilde{Z}_2$ defined in (\ref{deltagamma}). One has
\bea
\!\!\!\!\!\!\!\!\!\!d\widetilde{b}_0 \!\!&=&\!\! - *_4 d b_2 = -*_4 (\mathbf{v}_{uli}\,\partial_k \partial_l \mathcal{I}_j\,dx^i\wedge dx^j\wedge dx^k)=- \epsilon_{mijk}\,\mathbf{v}_{uli}\,\partial_k \partial_l \mathcal{I}_j\,dx^m\nonumber\\
\!\!&=&\!\!\frac{1}{2}\,\epsilon_{mijk}\,\epsilon_{lipq}\,\mathbf{v}_{upq}\,\partial_k \partial_l \mathcal{I}_j\,dx^m\nonumber\\
\!\!&=&\!\!\mathbf{v}_{ujk}\,\partial_k \partial_l \mathcal{I}_j\,dx^l+\mathbf{v}_{ukl}\,\partial_k \partial_j\mathcal{I}_j \,dx^l+\mathbf{v}_{ulj}\,\partial_k \partial_k \mathcal{I}_j\,dx^l\nonumber\\
\!\!&=&\!\! -d (\mathbf{v}_{ulk}\,\partial_l \mathcal{I}_k)\quad \Rightarrow\quad \widetilde{b}_0 = -\mathbf{v}_{ulk}\,\partial_l \mathcal{I}_k\,,
\eea
where we have used the anti-self-duality of $\mathbf{v}_{uij}$ and the properties (\ref{integralprop}) of $\mathcal{I}_i$. Similarly one finds that
\be
d\widetilde{Z}_2 = - *_4 d \widetilde{\gamma}_2 \quad \Rightarrow\quad \widetilde Z_2 = 1 + Q_5\,\mathcal{I} + \mathbf{v}_{ulk}\,\partial_l \mathcal{I}_k\,,
\ee
where we have picked the solution for $\widetilde{Z}_2$ that goes to 1 at infinity, and have used, to derive the $Q_5$-proportional term, the definition (\ref{Ihat}) of $\widehat{\mathcal{I}}_{ij}$.

We now have all the ingredients to verify the supergravity constraints. The equalities $\widetilde{Z}_1= Z_1$,  $\widetilde{Z}_2= Z_2$, $D=\frac{Z_1}{Z_2}$, $b_0 = \widetilde{b}_0 = c=f$, $\widetilde{a_1}=a_1$,  $\widetilde{b_1}=b_1$ are evidently satisfied by the metric coefficients listed above. The self-duality condition $da_3 = *_4 d a_3$ follows from anti-self-duality of $\mathbf{v}_{uij}$ and the fact that $\mathcal{I}$ is harmonic (\ref{integralprop}):
\bea
*_4 d a_3 &=& *_4(\mathbf{v}_{uli}\,\partial_j\partial_l \mathcal{I}\,dx^j\wedge dx^i) = \frac{1}{2}\,\epsilon_{kmji}\,\mathbf{v}_{uli}\,\partial_j \partial_l \mathcal{I}\,dx^k\wedge dx^m \nonumber\\
&=&-\frac{1}{4}\,\epsilon_{kmji}\,\epsilon_{lipq}\,\mathbf{v}_{upq}\,\partial_j \partial_l \mathcal{I}\,dx^k\wedge dx^m\nonumber\\
&=& -\frac{1}{2} (2\,\mathbf{v}_{umj}\,\partial_j \partial_k \mathcal{I}\,dx^k\wedge dx^m +\mathbf{v}_{ukm}\,\partial_j \partial_j \mathcal{I}\,dx^k\wedge dx^m)\nonumber\\
& =& da_3\,.
\eea
The proof of the self-duality of $db_1$ is identical, with the
replacement $\mathcal{I}\to \widetilde{\mathcal{I}}$. The same
identity, plus the definition of $\widehat{\mathcal{I}}_i$
(\ref{Ihat}), shows that $da_1$ is self-dual. The fact that $Z_1$,
$Z_2$, $Z_3$, $b_0$ and $k$ are harmonic follows from the harmonicity
of $\mathcal{I}$, $\widetilde{\mathcal{I}}$, and $\mathcal{I}_i$
(\ref{integralprop}).  We are left to show that the 4D metric $ds^2_4$
given in (\ref{basemetric}) is hyper-Kahler: this also follows from
anti-self-duality of $\mathbf{v}_{uij}$ and the properties of
$\mathcal{I}_i$, and we give the details of the proof in Appendix
\ref{sec:proofhyper}.

The supergravity ansatz of Section~\ref{sec:ansatz} reduces to the class
of supergravity solutions that have been used in the literature~\cite{Bena:2007kg} to describe black hole microstates when $b_0=\widetilde{b}_0=c=f=0$ and $b_1=\widetilde{b}_1=0$. We have seen that the string amplitude computation predicts that the class of D-brane configurations with 
equal D1 and D5 profiles emits non-zero values of these latter fields, and thus cannot be described by the existing microstate geometries. The fields $b_1$ and $b_0$ (that first appear at order $1/r^3$ and $1/r^4$, respectively) represent new types of dipole and quadrupole moments, proportional to both the D1-D5 vev $\mathbf{v}_{uij}$ and the derivative of the string profile $\dot{f}_i$, and thus vanish when any one of the three charges vanishes. This is in contrast with the three types of dipole moments of the existing microstate solutions, each of which survives in one of the three 2-charge limits. Since the new multipole moments involve all three charges, it is difficult to use dualities to relate them to a simpler system, as  can be done for the moments involving only two charges at a time. Another interesting outcome of our calculation is that it predicts that the 4D base metric $ds^2_4$, which is simply the flat metric on $\mathbb{R}^4$ in the 2-charge case, is a {\em non-trivial} hyper-Kahler metric when all three charges are non-vanishing. The non-flatness of the base metric for 3-charge microstate geometries was  already noted in the particular solution of \cite{Giusto:2004kj}, and it had remained until now a largely unexplained phenomenon. It is nice to see that 
our approach neatly predicts this feature.

\section{Discussion} \label{sec:discussion}

In this paper we showed how to extract information on the geometrical
backreaction of D-brane bound states, in the regime of finite
gravitational coupling, from perturbative string amplitudes. The
string amplitudes of interest involve both open and closed strings;
the open strings determine the state of the D-brane configuration and
the closed strings specify the supergravity field under
consideration. The most interesting contributions come from disk
amplitudes that mix different types of boundary conditions, in a
spirit very similar to the stringy description of classical gauge
instantons of~\cite{Billo:2002hm}. In this setup, the open strings
stretched between the instantonic and the physical branes are part of
the instanton moduli and so the physical observables are obtained
after integrating over these fields. For instance,
recently~\cite{Billo:2011uc} considered a ${\cal N}=2$ superconformal
setup and derived the backreaction on the axion-dilaton field due to
the presence of D$(-1)$-branes. As seen in the two-charge
cases~\cite{Giusto:2009qq,Black:2010uq}, in our construction the open
string vevs contain the data specifying the microstate and no
integration over the open string fields is necessary. We saw that also
the gravitational couplings of three-charge microstates are determined
by the open string data, which in our case are encoded by the
functions $f^i({\hat v})$ and the condensate $v_{IJK}$. Once these
couplings are derived from string theory, the leading gravitational
backreaction is obtained by solving the free bulk equations of
motion. As a consistency check, we also showed explicitly that the
bulk configurations derived in this way are consistent with the type
IIB equations of motion and preserve four supersymmetries, at least up
to fourth order in the $1/r$ expansion.

The results presented in this paper focus on a particular class of
three-charge bound states which has a simple world-sheet description,
as described in Section~\ref{epro}. Most likely a typical microstate of the
D1-D5-P system will not be in this class of configuration. In
addition, our result about the large distance behaviour of the
supergravity fields are non-trivial only if the wave profile is slowly
varying and its moments, such as $\int_0^{L_T} \dot{f}^i f^j
d\hat{v}$, are sizable. Again this is certainly not the case for a
generic microstate, where the direction (in the $\mathbb{R}^4$) of the
modes of the profile will be randomly distributed. As usual
we hope to learn something about the backreaction of a typical
microstate, even if we start by focusing on an atypical case
described by semiclassical data such as the profile functions $f^i$.

It is interesting to notice that the simplicity at the microscopic
level is not reflected in a particularly compact supergravity
solution. On the contrary, the geometric backreaction for a D1 and
D5-brane bound state with equal oscillations contains new types of
multipole moments that do not appear in
the class of $1/8$-BPS solutions studied in~\cite{Bena:2004de}. 
In discussing the regime of validity of our perturbative string calculation, 
we have given our reasons for believing that this more general type of asymptotic behavior applies also 
to states deep in the Higgs branch and for large $g_s N$.

It would be of course interesting to see whether we can engineer a D-brane configuration
which emits {\em only} the fields excited in the ansatz
of~\cite{Bena:2004de}. In our case, this would require to switch off
all fields in Section~\ref{sec:backreaction} proportional to
$\mathbf{v}_{ujl} \dot{f}^j \partial_l (1/r^2)$. For instance this
would happen if the profile function $f^i$ were made of two
disconnected circles in the $(x_1,x_2)$ and $(x_3,x_4)$ planes, but
this is not an allowed configuration for a microstate. It does not
seem to be simple to satisfy this requirement with an allowed
profile. This also means that the microstate
solutions~\cite{Giusto:2004id} are not described by our D-brane
configurations.

By extrapolating our solution to the case where the profile functions
on the D1 and D5 branes are different, one may write an educated guess
for the structure of a configuration where $f^{\rm D1}$ is a circle in
the $(x_1,x_2)$ plane and $f^{\rm D5}$ is given by the same circle but
now in the $(x_3,x_4)$ plane. It is possible that this is the D-brane
configuration whose backreaction reduces to the ansatz
in~\cite{Bena:2004de}. However, in order to analyze explicitly this
case, we need first to derive the possible states for an open string
stretched between D1 and D5 branes with different profiles.

Another interesting future line of development is to keep focusing on the class of configurations analyzed in this paper and derive a full non-linear ansatz solving the type IIB supersymmetry variations and equations of motion. 
The final goal would be to extend the relations
(\ref{functionstart})-(\ref{functionend}) to all orders in the
condensate $v_{IJK}$. The recent proposal of \cite{Bena:2011uw}, that
associates three-charge bound state configurations with functions of
two variables, would suggest that the all-order form of the expressions 
(\ref{functionstart})-(\ref{functionend}) could be
represented in terms of integrals of the type appearing in
Eq.~(\ref{integraldef}), but with the profile $f^i(\hat v)$ replaced
by a function of two variables. If this program could be completed, it
would represent a major development towards the construction of a
family of geometries with enough degrees of freedom to encode for the
full entropy of the three-charge black hole.

A more immediate step towards this goal would be to focus on the subclass of configurations with two axial symmetries. Work in progress indicates that exact solutions within this class can be constructed and it would be very interesting to study the simplest explicit solution of this ansatz.  This configuration could play the role the solution
in~\cite{Giusto:2004id} played for the ansatz~\cite{Bena:2004de}. An analysis
of the ``near-horizon'' limit of such a solution has the potential to
provide, via the AdS/CFT correspondence, further evidence that we are
really considering the geometrical backreaction of a three-charge
microstate.

\vspace{0.6cm}

\noindent {\large \textbf{Acknowledgements} }

\vspace{2mm}
We thank I.~Bena, W.~Black, G.~Dall'Agata, S.~El-Showk, V.~Jejjala, S.~Mathur, J.F.~Morales, S.~Ramgoolam, C.~Ruef, A.~Strominger, B.~Vercnocke, N.~Warner, E.~Witten for discussions.
The work of DT at QMUL was supported by an STFC studentship.

\appendix 

\section{Constraints from supersymmetry} \label{sec:susy_details}

\subsection{Killing spinor equations}

In our conventions the supersymmetry variations of the gravitino and dilatino in IIB theory, in units where $\kappa=1$, are
\begin{eqnarray}
\label{eq:gravitino}
	\delta \psi_M & = &  \left( \nabla_M - \frac{i}{2}Q_M\right)\epsilon + \frac{i}{192} \, \Gamma^{M_1 \ldots M_4} F^{(5)}_{M_1\ldots  M_4 M} \epsilon 
\\ \nonumber & -& \frac{1}{96} \,  G_{NPQ} \Gamma_M^{~~NPQ} \epsilon^* + \frac{9}{96} \, G_{MNP} \Gamma^{NP} \epsilon^* \,, \\
	\delta \lambda & = & i \Gamma^M P_M \epsilon^* + \frac{i}{24}  G_{MNP} \Gamma^{MNP} \epsilon\,, \label{eq:dilatino} 
\end{eqnarray}
where
\begin{equation}
	\begin{array}{rcl}
		P &=& \displaystyle \frac{i}{2}{\rm e}^{\phi} dC^{(0)} +\frac12 d \phi ,\\[2mm]
		Q &=& \displaystyle -\frac12 {\rm e}^{\phi} dC^{(0)},\\[2mm]
		G &=& \displaystyle i {\rm e}^{\phi/2}\left(\tau dB - dC^{(2)}\right), \\[2mm]
		\tau&=& C^{(0)}+ i e^{-\phi}\,.
	\end{array}
	\label{collegamento} 
\end{equation}
We take the supersymmetry parameter $\epsilon$ to satisfy the chirality condition
\be
\Gamma^{0y12345678}\,\epsilon = \epsilon\,,
\ee
where $1,2,3,4$ are the directions of $\mathbb{R}^4$ and $5,6,7,8$ the $T^4$ directions. Correspondingly $F^{(5)}$ satisfies $F^{(5)}= * F^{(5)}$, where the star operation is defined using the orientation $\epsilon_{0y12345678}=1$. We are using a base in which the 10D gamma matrices are purely imaginary, in which case the conjugate $\epsilon^*$ of $\epsilon$ is simply given by complex conjugation: $\epsilon^* = \epsilon_1 - i \epsilon_2$, with $\epsilon= \epsilon_1 + i \epsilon_2$ and $\epsilon_1$, $\epsilon_2$ real spinors. We will denote the $\mathbb{R}^4$ coordinates by $i,j,\ldots = 1,2,3,4$ and the $T^4$ coordinates by $a,b,\ldots = 5,6,7,8$.

\subsection{Vielbeins, spin connection and gauge fields}
To explicitly write the Killing spinor equations one needs the vielbeins and spin connection of the Einstein frame metric ($ds^2_E = e^{-\phi/2} ds^2$) and the gauge fields for the general ansatz specified in section \ref{sec:ansatz}. We give these data below, keeping only the terms that contribute to the large distance expansion up to order $1/r^4$. 

The vielbeins are
\bea
&&\!\!\!\!\!\!\!\!\!\!\!\!\!\!\!\!\ e^t =\frac{1}{(Z_1 Z_2)^{1/4} Z_3^{1/2} D^{1/8}} \,(dt+k)\,,\quad e^y = \frac{Z_3^{1/2}}{(Z_1 Z_2)^{1/4} D^{1/8}} \Bigl(dy+dt-\frac{dt+k}{Z_3}+a_3\Bigr)\,,\nonumber\\
&&\!\!\!\!\!\!\!\!\!\!\!\!\!\!\!\!\ e^i = \frac{(Z_1 Z_2)^{1/4}}{D^{1/8}} \eo^i\,,\quad e^a = \Bigl(\frac{Z_1}{Z_2}\Bigr)^{1/4}\frac{1}{D^{1/8}} dx^a\,,
\eea
with $\eo^i$ the vielbeins of the metric $ds^2_4$.
The non-trivial components of the spin connection are
\bea
\omega_{ti}&=&\frac{D^{1/8}}{(Z_1 Z_2)^{1/4}}\Bigl(\frac{1}{4} \partial_i \log Z_1+\frac{1}{4} \partial_i \log Z_2+\frac{1}{2} \partial_i \log Z_3+\frac{1}{8}\partial_i \log D \Bigr)\,e^t\nonumber\\
&&+\frac{1}{2}\frac{D^{1/8}}{(Z_1 Z_2)^{1/4}}\partial_i \log Z_3 \,e^y-\frac{1}{2} (\partial_i k_j -\partial_j k_i)\,e^j\,,\cr
\omega_{ty}&=&\frac{1}{2} \frac{D^{1/8}}{(Z_1 Z_2)^{1/4}} \partial_i \log Z_3 \,e^i\,,\cr
\omega_{yi}&=&\frac{D^{1/8}}{(Z_1 Z_2)^{1/4}}\Bigl(-\frac{1}{4} \partial_i \log Z_1-\frac{1}{4} \partial_i \log Z_2+\frac{1}{2} \partial_i \log Z_3-\frac{1}{8} \partial_i \log D\Bigr)\,e^y\nonumber\cr
&&+\frac{1}{2}\frac{D^{1/8}}{(Z_1 Z_2)^{1/4}}\partial_i \log Z_3 \,e^t+\frac{1}{2} (\partial_i a_{3j}-\partial_j a_{3i}-\partial_i k_j +\partial_j k_i)\,e^j\,,\nonumber\cr
\\
\omega_{ij}&=&\frac{D^{1/8}}{(Z_1 Z_2)^{1/4}}\Bigl(\frac{1}{4} \partial_j \log Z_1+\frac{1}{4} \partial_j \log Z_2-\frac{1}{8}\partial_j \log D\Bigr)\,e^i-(i\leftrightarrow j)\nonumber\cr
&&+\frac{1}{2}  (\partial_i k_j -\partial_j k_i)\,e^t - \frac{1}{2} (\partial_i a_{3j}-\partial_j a_{3i}-\partial_i k_j +\partial_j k_i)\,e^y + \overline{\omega}_{ij}\,,\\
\omega_{ai}&=&\frac{D^{1/8}}{(Z_1 Z_2)^{1/4}}\Bigl(\frac{1}{4} \partial_i \log Z_1 -\frac{1}{4} \partial_i \log Z_2 -\frac{1}{8} \partial_i \log D\Bigr)\,e^a\,,
\eea
where $\overline{\omega}_{ij}$ is the spin connection of $ds^2_4$.

The gauge fields are
\be
P = \frac{i}{2} \partial_i c\,e^i + \frac{1}{4}\frac{D^{1/8}}{(Z_1 Z_2)^{1/4}}\,\partial_i \log D\,e^i\,,
\ee
\be
Q =  -\frac{1}{2}\partial_i c\,e^i\,,
\ee
\bea
G 
&=& - i \Bigl(i \partial_i b_0 + D^{5/8} (Z_1 Z_2)^{1/4}\,\frac{\partial_i \widetilde Z_1}{\widetilde Z_1^2}\Bigr)\,e^i\wedge e^t\wedge e^y \cr
&&- i( -i \partial_i b_{1 j} +\partial_i a_{1 j}-\partial_i k_j)\,e^i\wedge e^j\wedge e^y \cr
&&-i(-i\partial_i \widetilde b_{1j}+ \partial_i \widetilde a_{1 j}+\partial_i a_{3 j} - \partial_i k_j)\,e^i\wedge e^j\wedge e^t \cr
&&-i\,\frac{\epsilon_{ijkl}}{3!}\, \Bigl(-i\partial_l \widetilde b_0 +\frac{D^{5/8}}{(Z_1 Z_2)^{3/4}}\,\partial_l \widetilde Z_2\Bigr)\,e^i\wedge e^j\wedge e^k\,.
\eea

We will analyze below the constraints coming from imposing $\delta \psi_M =\delta \lambda=0$ order by order in the $1/r$ expansion.

\subsection{Order $1/r^2$}
At order $1/r^2$ the only non-trivial functions are $Z_1=\widetilde Z_1, Z_2=\widetilde Z_2, Z_3$ and $D=Z_1/Z_2$.\footnote{By reversing the sign of the RR fields, one could have also taken 
$\widetilde Z_1= -Z_1$ and $\widetilde Z_2 = -Z_2$. Sending all the RR fields to minus themselves and $\epsilon\to \epsilon^*$ leaves the supersymmetry variations (\ref{eq:gravitino}), (\ref{eq:dilatino}) invariant and constitutes a symmetry of the theory. Hence our choice is not restrictive.}
At this order the dilatino equation $\delta \lambda=0$ becomes
\be
i (\partial_i Z_1 - \partial_i Z_2)\,\Gamma^i \,\epsilon^* + \partial_i Z_1\,\Gamma^{ity}\,\epsilon + \frac{1}{3!} \epsilon_{ijkl} \,\partial_l Z_2\,\Gamma^{ijk}\,\epsilon=0\,.
\ee
Requiring the coefficients of $\partial_i Z_1$ and $\partial_i Z_2$ to vanish separately gives
\be\label{susyconstraint1}
\Gamma^{ty}\,\epsilon_1=-\epsilon_2  \,,\quad \Gamma^{1234} \,\epsilon_1 = - \epsilon_2  \,. 
\ee 
No new constraint is imposed by the $M=a$ components of the gravitino equation $\delta \psi_M=0$. The $M=t$ component of the gravitino equation is
\be
\Bigl(\frac{3}{4}\partial_i Z_1 + \frac{1}{4}\partial_i Z_2 + \partial_i Z_3 \Bigr) \Gamma^{ti}\,\epsilon + \partial_i Z_3 \,\Gamma^{y i}\epsilon - \frac{i}{4} \frac{1}{3!} \epsilon_{jkli} \,\partial_i Z_2\,{\Gamma}^{tjkl} \epsilon^*+i \frac{3}{4} \partial_i Z_1\, \Gamma^{iy} \,\epsilon^*=0\,,
\ee
and one has an equivalent equation from $M=y$. The $Z_1$ and $Z_2$ terms vanish thanks to 
(\ref{susyconstraint1}); the $Z_3$ term implies:
\be
\Gamma^{ty}\epsilon_1 = \epsilon_1\,,
\ee
which together with (\ref{susyconstraint1}) gives
\be
\epsilon_2 = - \epsilon_1 \,.
\ee
Finally the $M=i$ components of the gravitino equation yield
\bea
&&\partial_i \epsilon + \frac{1}{4} \partial_i Z_3 \,\Gamma^{ty}\epsilon + \frac{1}{16}\Bigl(\partial_j Z_1 + 3 \partial_j Z_2  \Bigr) \Gamma^{ij}\,\epsilon +\frac{i}{16} (\partial_j Z_1 \Gamma^{ijty} \epsilon^* - \partial_i Z_2 \Gamma^{1234} \epsilon^*)\nonumber\\
&&\qquad \qquad-i  \frac{3}{16}\Bigl(\partial_i Z_1 \Gamma^{ty}\epsilon^* + \frac{1}{2} \epsilon_{ijkl} \partial_j Z_2 \,\Gamma^{kl}\,\epsilon^*\Bigr)=0\,.
\eea
Using the constraints derived above, this equation reduces to a differential equation for $\epsilon_1$, which is solved by
\be\label{epsilon1byr2}
\epsilon_1 = Z_1^{-3/16}\,Z_2^{-1/16}\,Z_3^{-1/4}\,\epsilon_0\,,
\ee
with $\epsilon_0$ a constant spinor. 

In summary the spinor satisfies the projection conditions
\be\label{susyprojections}
\Gamma^{ty}\epsilon_1 = \epsilon_1\,,\quad \Gamma^{1234}\epsilon_1 = \epsilon_1\,,\quad\Gamma^{5678}\epsilon_1=\epsilon_1\,,\quad \epsilon_2= - \epsilon_1\,,
\ee
where the third constraint follows from the previous ones and the chirality condition.
These constraints leave 4 independent components, corresponding to the supersymmetries preserved by a 3-charge black hole. We can use these projection conditions in the computation at order $1/r^3$.
\subsection{Order $1/r^3$}
At order $1/r^3$ the only new constraints coming from supersymmetry are the ones involving the 1-forms. We will analyze these new conditions in the following. 

The real and imaginary parts of the dilatino equation, or the $M=a$ components of the gravitino equation, imply, after using (\ref{susyprojections}):
\be
(\partial_i a_{3 j} + \partial_i \widetilde a_{1j}-\partial_i a_{1j})  \Gamma^{ij}\epsilon_1=0\,,
\ee
and
\be
 (\partial_i \widetilde b_{1j}-\partial_i b_{1j}) \Gamma^{ij}\epsilon_1=0\,.
\ee
Since the condition $\Gamma^{1234}\epsilon_1=\epsilon_1$ implies that
\be
\Gamma^{ij}\epsilon_1 =  - \frac{1}{2} \epsilon_{ijkl}\, \Gamma^{kl}\epsilon_1
\ee
an equation of the form $\omega_{ij}\,\Gamma^{ij}\epsilon_1=0$, for some 2-form $\omega_{ij}$, requires that the anti-self-dual part of $\omega_{ij}$ vanish, i.e. that  $\omega = *_4 \omega$. Hence the two conditions above are equivalent to
\be
(1-*_4)(da_3 + d\widetilde a_1 - d a_1)=0\,,\quad (1-*_4) (db_1- d\widetilde b_1)=0\,.
\ee
In an analogous way, the $M=t$ and $M=y$ components of the gravitino equation imply
\be
(1-*_4)(d a_{1} +3 d a_{3} +3 d \widetilde a_{1})=0\,,\quad (1-*_4)(d b_{1} + 3 d \widetilde b_{1})=0\,,
\ee
and
\be
(1-*_4)(d \widetilde a_{1} -3d a_{3} +3 d a_{1} )=0 \,,\quad (1-*_4)(d \widetilde b_{1} + 3 d b_{1})=0\,.
\ee
Altogether these conditions require that all the 1-forms (apart from $k$)  have self-dual field strengths:
\be
(1-*_4) da_1 = (1-*_4) d\widetilde a_1=(1-*_4) da_3=(1-*_4) db_1=(1-*_4) d\widetilde b_1\,.
\ee
Let us now consider the $M=i$ components of the gravitino equation: the terms involving the scalars give the same differential equation for $\epsilon$ found at order $1/r^2$, and hence $\epsilon$ is given by an expression of the form \ref{epsilon1byr2} even at $1/r^3$ order. The terms involving the 1-forms can be simplified by the use of the identity
\be
\omega_{jk}\,\Gamma^{ijk}\epsilon_1 = - 2 \,\omega_{ij}\,\Gamma^{j}\epsilon_1\,,
\ee
valid for any self-dual 2-form $\omega_{ij}$ if $\Gamma^{1234}\epsilon_1 = \epsilon_1$. Then the real and imaginary parts of the $M=i$ gravitino equation give
\be
( \partial_{[i} \widetilde a_{1 j]}-\partial_{[i} a_{1j]} )\,\Gamma^{j}\epsilon_1=0\,,
\ee
and
\be
(\partial_{[i} \widetilde b_{1j]}- \partial_{[i} b_{1j]})\,\Gamma^{j}\epsilon_1=0\,,
\ee
which imply
\be
a_1 = \widetilde a_1 \,,\quad b_1 = \widetilde b_1\,.
\ee

\subsection{Order $1/r^4$}
The equations for the 1-forms are unchanged at order $1/r^4$, therefore we will only discuss the scalar and 4D metric sector below. 
 
 The conditions following from the real and imaginary part of the dilatino equation are
 \be
\partial_i \log D - D^{1/2}\,\Bigl((Z_1 Z_2)^{1/2}\frac{\partial_i \widetilde Z_1}{\widetilde Z_1^2}-(Z_1 Z_2)^{-1/2}\partial_i \widetilde Z_2   \Bigr) =0\,,
\ee
\be
2 \,\partial_i c -  \partial_i b_0- \partial_i \widetilde b_0=0\,,
\ee
and they imply, using the asymptotic conditions (\ref{asymptoticbc}), that
\be
D=\frac{\widetilde Z_1}{\widetilde Z_2}\,,\quad c = \frac{1}{2} (b_0 + \widetilde b_0)\,.
\ee
The $M=a$ gravitino equation gives
\be
2\,\partial_i \log\frac{Z_1}{Z_2}  - \partial_i \log D -D^{1/2} \Bigl((Z_1 Z_2)^{1/2}\frac{\partial_i \widetilde Z_1}{\widetilde Z_1^2}-(Z_1 Z_2)^{-1/2}\partial_i \widetilde Z_2 \Bigr) =0\,,
\ee
\be
2\,\partial_i f - \partial_i b_0-  \partial_i \widetilde b_0=0\,.
\ee
Combining these conditions with the previous ones gives the further constraints
\be
\frac{\widetilde Z_1}{\widetilde Z_2}=\frac{Z_1}{Z_2}\,,\quad f = c= \frac{1}{2} (b_0 + \widetilde b_0)\,.
\ee
The conditions following from the $M=t$ gravitino equation are
\be
2\,\partial_i \log (Z_1 Z_2)+\partial_i \log D - D^{1/2} \Bigl(3\, (Z_1 Z_2)^{1/2}\,\frac{\partial_i \widetilde Z_1}{\widetilde Z_1^2}+ (Z_1 Z_2)^{-1/2}\,\partial_i \widetilde Z_2\Bigr)=0\,,
\ee
\be
2\,\partial_i f + \partial_i \widetilde b_0-3\,  \partial_i b_0=0\,,
\ee
and imply, together with the previous conditions,
\be
\widetilde Z_1 \widetilde Z_2=Z_1 Z_2  \,,\quad \widetilde b_0 = b_0\,.
\ee
The $M=y$ gravitino equation introduces no new constraints.

In summary one has
\be\label{susy4th}
\widetilde Z_1=Z_1\,,\quad \widetilde Z_2 = Z_2 \,,\quad \widetilde b_0 = b_0\,,\quad c=f=b_0\,.
\ee

In the $M=i$ components of the gravitino equation the terms in $\epsilon$ of order $1/r^4$  contribute, and one thus has to consider the possibility that the relation $\epsilon_1 = -\epsilon_2$ be violated by order $1/r^4$ terms. Hence one can write
\be
\epsilon= (1-i)\epsilon_1 + i \widetilde \epsilon_2\,,
\ee
where $\widetilde\epsilon_2 = O(r^{-4})$.  The equations one gets after taking into account the identities 
(\ref{susy4th}) are
\be
\frac{D^{1/8}}{(Z_1 Z_2)^{1/4}} \Bigl[\partial_i \epsilon_1+\partial_i \log (Z_1^{3/16} Z_2^{1/16} Z_3^{1/4})\epsilon_1\Bigr] -\frac{1}{2}\partial_i \widetilde\epsilon_2+\frac{1}{4} \overline{\omega}_{jk,i}\Gamma^{jk}\epsilon_1=0\,,\label{eqmi1}
\ee
and
\be
\partial_i \widetilde\epsilon_2 + \partial_i b_0\, \epsilon_1=0\,.
\label{eqmi2}
\ee
The second equation simply determines $\widetilde\epsilon_2$ to be
\be
\widetilde\epsilon_2 = -b_0\, \epsilon_0\,.
\ee
The first equation determines $\epsilon_1$:
\be
\epsilon_1 = Z_1^{-3/16} Z_2^{-1/16} Z_3^{-1/4}\,\Bigl(1-\frac{1}{2} b_0\Bigr)\,\epsilon_0 + \widetilde \epsilon_1 \,,
\ee
with
\be
\partial_i \widetilde \epsilon_1 + \frac{1}{4}\, \overline{\omega}_{jk,i}\Gamma^{jk}\epsilon_0 = 0\,.
\ee
The compatibility condition for the equation above is
\be
\partial_{[l}  \,\overline{\omega}_{jk,i]}\Gamma^{jk}\epsilon_0=\overline{R}_{jk,li}\,\Gamma^{jk}\epsilon_0 + O(r^{-5})=0\,,
\ee
where $\overline R_{ij,kl}$ is the curvature of $ds^2_4$. Remembering that $\Gamma^{1234}\epsilon_0 = \epsilon_0$, the compatibility equation is equivalent to
\be
\overline{R}_{ij,kl} = \frac{1}{2}\,\epsilon_{ijrs}\,\overline{R}_{rs,kl}\,,
\ee
i.e. the metric $ds^2_4$ is hyper-Kahler.

\section{Closed string vertices} \label{app:vertices}

In this appendix we summarize our conventions for the world-sheet CFT
and discuss some details of the closed string vertices used in the
disk amplitudes of Section~\ref{sec:backreaction}. The holomorphic components of the
string fields satisfy the standard OPE relations
\begin{eqnarray}
\partial X^\mu(z) \partial X^\nu(w) \sim 
-\frac{2\alpha'\eta^{\mu\nu}}{(z-w)^2}~&,&~~~
c(z) b(w) \sim \frac{1}{z-w}~~,~~~~
\\ \nonumber 
\psi^\mu(z) \psi^\nu(w) \sim \frac{\eta^{\mu\nu}}{z-w}~&,&~~~
\gamma(z) \beta(w) \sim \frac{1}{z-w}~,
\end{eqnarray}
where $b,c$ ($\beta,\gamma$) are the usual (super)ghost fields, and
the full closed string coordinate $x^\mu$ is given by
$x^\mu(z,\bar{z}) = (X^\mu(z)+X^\mu(\bar{z}))/2$. The simplest form
for the vertex operator describing the emission of a massless NS-NS
string state is
\begin{equation}
W_{NSNS}^{(-1,-1)} =  {\cal G}_{\mu\nu}
 \psi^{\mu} \ex{-\varphi} \,
\widetilde\psi^{\nu} \ex{-\widetilde\varphi} 
\ex{ \ii  k \cdot x}~,
\label{clNS-1} 
\end{equation}
where as usual the bosonic field $\varphi(z)$ with background charge
$-2$ and the fermionic system $(\eta,\xi)$ of conformal weight
$(1,0)$ provide an equivalent description for the superghost sector
\begin{equation}
  \label{bosonization}
  \gamma \simeq \ex{\varphi} \eta~~,~~~~
  \beta \simeq \partial \xi \ex{-\varphi}~.
\end{equation}
The BRST charge is (we follow the conventions of~\cite{Polchinski:1998rr})
\begin{equation} \label{BRST}
Q_{\rm B}  = \oint \frac{dz}{2 \pi i} \left\{ \;\! c \;\! \Big( 
T_X + T_\psi + T_{\beta,\gamma} + (\partial c)b\Big) 
+\gamma \;\! j_{X,\psi} - b \;\! \gamma^2 \right\}~,
\end{equation}
where the (holomorphic parts of the) stress energy tensor and
supercurrent are
\begin{eqnarray}\label{Te}
T_X(z) & = & -\frac{1}{4\alpha'} \partial X^\mu \partial X_\mu~,~~~~
T_\psi(z) = - \frac{1}{2} \psi^\mu \partial \psi_\mu~, 
\\ \label{Sue}
T_{\beta,\gamma}(z) &=& \frac{1}{2} (\partial \beta) \gamma - 
\frac{3}{2} \beta \partial\gamma~,~~
j_{X,\psi}(z) =  \frac{i}{\sqrt{2\alpha'}} \psi^\mu \partial X_\mu~.
\end{eqnarray}
The vertex operators~\eqref{clNS0} and~\eqref{clNS-1} are invariant
under separate holomorphic and antiholomorphic BRST variations,
provided that we restrict to massless ($k^2=0$) and transverse ($k^\mu
{\cal G}_{\mu\nu}=0$) states.

While we can use the RR vertices~\eqref{clR0} directly in the string
amplitudes we are interested in, we need to discuss in more
detail the NSNS vertices. First we must separate the dilaton and
the graviton parts of the polarization ${\cal G}_{\mu\nu}$, then we
must find a representative for these state whose total (holomorphic
plus antiholomorphic) charge is $-1$, instead of $-2$. 

The standard way to separate the dilaton and the graviton terms is to
write the symmetric part of ${\cal G}$ in two parts, with the dilaton contribution
coming from the part proportional to
\begin{equation}
  \label{dilus}
\epsilon^{\rm dil}_{\mu\nu} =  \eta_{\mu\nu} - k_\mu \ell_\nu- k_\nu \ell_\mu~,
\end{equation}
with $\ell^2=0$ and $\ell_\mu k^\mu=1$,
and the graviton contribution coming from the terms orthogonal to~\eqref{dilus}.  
However this requires a non-zero value for the momentum $k$ and requires one to choose explicitly the light-cone by
fixing the null-vector $\ell$. A covariant way to separate graviton
and dilaton contributions is to choose different BRST representative
for their vertex operators. By following~\cite{Bergman:1994qq}, it is possible to
show that for $k^2=0$ the vertex operator
\begin{equation}
  \label{newdil}
W_{\rm dil}^{(-2)} = \left( \eta_{\mu\nu}
 \psi^{\mu} \ex{-\varphi} \widetilde\psi^{\nu} \ex{-\widetilde\varphi} 
+  c  \eta \widetilde{c} \, \bar\partial\widetilde\xi \ex{-2\widetilde\varphi} - 
 c \partial\xi  \ex{-2\varphi} \widetilde{c} \, \widetilde{\eta}\right) 
\ex{ \ii  k \cdot x}~
\end{equation}
is in the BRST-cohomology of $Q_{\rm B}+\widetilde{Q}_{\rm B}$\footnote{Notice
  that~\eqref{newdil} is not annihilated by $Q_{\rm B}$ and $\widetilde{Q}_{\rm B}$
  separately; also the picture of this state cannot be separated in
  its left and right moving part and is given by the sum of the
  eigenvalues of the operator $\oint \frac{dz}{2\pi i} (\xi
  \eta-\partial\varphi)$ and its anti-holomorphic analogue.} and for
$k\not=0$ is equivalent to the vertex~\eqref{clNS-1} with the dilaton
polarization~\eqref{dilus}. Even if the vertex operator~\eqref{newdil}
seems rather complicated, the state obtained via the usual
operator/state correspondence is less so:
\begin{equation}
  \label{stateop}
  \lim_{z\to 0} W_{\rm dil}^{(-2)} \ket{0} = \left(
\eta_{\mu\nu} c_1 \psi^\mu_{-\frac 12} \widetilde{c}_1 \widetilde{\psi}^\nu_{-\frac 12} +
c_1 \gamma_{-\frac 12} \widetilde{c}_1 \widetilde\beta_{-\frac 12} - 
c_1 \beta_{-\frac 12} \widetilde{c}_1 \widetilde\gamma_{-\frac 12}\right)
\ket{k}_{-1} \widetilde{\ket{k}}_{-1}\,
\end{equation}
where $\ket{0}$ is the $SL(2,C)$ invariant vacuum (annihilated by
$\gamma_r$ with $r>1/2$ and $\beta_s$ with $s>-3/2$), while the states
labelled with $-1$ are annihilated by all superghost oscillators with
$r,s \geq 1/2$. Notice that the state~\eqref{stateop} appears
generically in the expansion of the NSNS part of the full boundary
state (see for instance~\cite{Billo:1998vr}), supporting the claim that this is
the form of the dilaton state to be used in disk amplitudes without
momentum flow in the Neumann directions.
 
For our purposes, we need to raise the picture of the
vertex~\eqref{newdil}. Again we cannot treat the holomorphic and the
anti-holomorphic part separately and thus we have to calculate 
$W_{NSNS}^{(-1)} = \left\{ Q_{\rm B} + \widetilde{Q}_{\rm B} ,(\xi + \widetilde{\xi}) W_{NSNS}^{(-2)} \right\} $, 
which yields
\begin{eqnarray} \nonumber
  W_{NSNS}^{(-1)} & = & \eta_{\mu\nu}
  \left[(\partial X^\mu - i \alpha' \, k \!\cdot\! \psi \, \psi^\mu) \;\!
  \widetilde{\psi}^\nu\ex{-\widetilde{\varphi}} + \psi^\mu\ex{-\varphi}
  (\bar\partial X^\nu - i \alpha' \, k \!\cdot\! \widetilde\psi \,
  \widetilde\psi^{\nu}) \right]   
  c \widetilde{c} \, \ex{i k \cdot x} \\   \label{newdil-1} 
  && {} +
  \sqrt{\frac{\alpha'}{2}} \left[k \!\cdot\! \partial(\psi \ex{\varphi}) c
    \eta \widetilde{c} \, \bar\partial\widetilde{\xi}\ex{-2\widetilde{\varphi}} -
     c \partial{\xi}\ex{-2\varphi} k \!\cdot\! \bar\partial(\widetilde{\psi}
     \ex{\widetilde{\varphi}}) \widetilde{c} \, \widetilde{\eta}\right] \ex{i k \cdot x}~. 
\end{eqnarray}
	
Let us consider what happens when this vertex is inserted in the
amplitude~\eqref{c}. Because of the structure of the open string
condensate discussed in Section~\ref{tos}, the only non-trivial
contributions to this amplitude come from the terms in the correlator
that, after the identification of the left/right moving fields,
contain three $\psi$'s. Thus we can drop the second line as it is at
most linear in $\psi$ and focus on the terms in the first line
of~\eqref{newdil-1}. The first of such terms was discussed in
Section~\ref{sec:2_boundaries}, so now we want to show that the second
term, where the holomorphic part is in the $-1$ picture and the
antiholomorphic one in the zero picture, yields exactly the same
result.

The calculation of this term differs from the one discussed in
Section~\ref{sec:2_boundaries} in two respects: first we have to
identify two anti-holomorphic fermionic fields and so we clearly have
contributions that are quadratic in the reflections
matrix~\eqref{eq:D1_bcs} or~\eqref{eq:D5_bcs} (again we can use either
of these two matrices, as they are identical in the $\mathbb{R}^{1,5}$
which is relevant for our purposes); then we have also to consider the
non-linear nature of the bosonic boundary conditions~\eqref{nln}. We
will show that the extra contributions which are related to these two
new features compensate each other. Actually this happens not just for
the terms in the dilaton vertex~\eqref{newdil-1}, but for a generic
NS-NS state in the $(-1,0)$ picture. Thus the net result for the
amplitude~\eqref{c} obtained from these NS-NS vertices is indeed
identical to the contribution obtained in
Section~\ref{sec:2_boundaries} with the $(0,-1)$ vertices.

A first way of obtaining three $\psi$'s in the correlator is to start
from the term $(\psi\, k \cdot \widetilde{\psi} \, \widetilde{\psi})$ and
apply the identification~\eqref{psii} twice. By
using~\eqref{eq:D1_bcs} or~\eqref{eq:D5_bcs}, we obtain from the
reflection matrix contracted with the momentum $k$ 
\begin{equation}\label{tppsi}
 k \!\cdot\! \widetilde{\psi} = - k \!\cdot\! {\psi} + 2 \;\! k \!\cdot\! \dot{f} \, \psi^v~.
\end{equation}
For the first term, one can follow exactly the same steps discussed in
Section~\ref{sec:2_boundaries} and obtain the result
in~\eqref{ampf0I-3}. Notice that the $-1$ present in the diagonal
terms $R^i_{\;i}$ is compensated by the different ordering of the
three fermionic fields. From the second term in~\eqref{tppsi} we get a
new contribution to the mixed disk amplitude which reads
\begin{equation}
  \label{psi3} i \alpha'
  \int\limits_0^{L_T} \!d\hat{v}\; 2 \;\! k \!\cdot\! \dot{f} \left({\cal G}
    R\right)_{ij} v^{ivj} e^{-ik \cdot f(\hat{v})} ~.
\end{equation}
If in this equation we take the $f$-independent part of the
identification matrix, then this is the integral, over a full period,
of the derivative of a periodic function $f(\hat{v})$. Then the only
non-trivial contribution from~\eqref{psi3} is from the $f$-dependent
component $\T{R}{u}{i}\,$,
\begin{equation} \label{psi4}
  \eqref{psi3} = -8i \alpha' {\cal G}_{iu}
  \int\limits_0^{L_T} \! d\hat{v} \,( k \!\cdot\! \dot{f} ) \dot{f}_j \;\!	
  v^{ivj} e^{-ik \cdot f(\hat{v})} ~. 
\end{equation}
Let us now focus on the contribution coming from $\bar\partial
X^u$. So far we neglected all terms of this type because they did not
give rise to correlators with the necessary three insertions of the
$\psi$-field. However this case is different and by using the
non-linear part of the identification in~\eqref{nln} we obtain
\begin{equation}
  \label{psi5}
  -8 \a'  {\cal G}_{iu} \int\limits_0^{L_T} \! d\hat{v}\, \ddot{f}_j v^{ijv}
  e^{-ik \cdot f(\hat{v})} =
  -8 i \a'  {\cal G}_{iu} \int\limits_0^{L_T} \! d\hat{v}\, ( k \!\cdot\! \dot{f} ) \dot{f}_j \;\!
  v^{ijv}  e^{-ik \cdot f(\hat{v})} ~,
\end{equation}
where we integrated by parts the double derivative $\ddot{f}$. This
result cancels~\eqref{psi4} and this completes the proof of the
equivalence between the vertices in $(-1,0)$ and $(0,-1)$ pictures.

\section{Spinor conventions} \label{app:conventions}

We use the spinor conventions of \cite{Giusto:2009qq}, which we record here
for completeness.

In our conventions, the 10D Majorana-Weyl spinors
$\Theta_{\hat A}$ satisfy $\Gamma_{(10)} \Theta_{\hat A}=-\Theta_{\hat
A}$, where $\Gamma_{(10)}=\Gamma^0_{(10)}\Gamma^y_{(10)} \Gamma^1_{(10)}
\ldots \Gamma^8_{(10)}$. These spinors decompose with respect to the
$SO(1,5)\times SO(4)$ as
\be
\Theta_{\hat A}= \{ \Theta_{A}^{~\dot\alpha} ; \Theta^{A \alpha} \} \,,
\ee
where upper and lower indices $A,B,\dots=1,\ldots,4$ denote Weyl
$SO(1,5)$ spinors of opposite chirality; similarly $\alpha,~
\dot{\alpha}=1,2$ are Weyl spinor indices of opposite chirality for
the $SO(4)$ group acting along the ND $T^4$ directions. We decompose
the 10D Gamma matrices as follows
\beq
\Gamma^a_{(10)} = 1_{(6)} \otimes \gamma^a~~,\qquad
\Gamma^I_{(10)} = \Gamma^I \otimes \gamma^{ND}~,
\eeq
where we use simply $\Gamma^I$ for the 6D Gamma matrices and
\bea
(\gamma^{ND})_{\dot\alpha}^{\dot\beta} &=&
(\prod_a \gamma^a)_{\dot\alpha}^{\dot\beta} = 
-\delta_{\dot\alpha}^{\dot\beta}
~,~~~  (\gamma^{ND})_{\alpha}^{\beta}=
(\prod_a \gamma^a)_{\alpha}^{\beta} = 
\delta_{\alpha}^{\beta} ~,\nn\\
(\Gamma)_{A}^{~B} &=&
(\prod_I \Gamma^I)_{A}^{~B}= 
-\delta_{A}^{~B}
~,~~~
(\Gamma)^{A}_{~B}=(\prod_I \Gamma^I)^{A}_{~B}=
\delta^{A}_{~B}\,.
\label{gGamma}
\eea
Instead of the 6D Gamma matrices, we will often use the chiral
components such as $ (C \Gamma^{I_1..I_{2n-1}})_{AB}$, where $C$ is the
6D charge conjugation matrix\footnote{$C$ is related to the 10D and 4D
charge conjugation matrices by $C_{10}=C\otimes C_4$.} satisfying ${}^{\rm
t}\Gamma^I= -C \Gamma^I C^{-1}$.

\section{Proof that the 4D base metric is hyper-Kahler}\label{sec:proofhyper}

The form of the base metric $ds^2_4$ predicted by the string theory
computation is 
\be
ds^2_4 = \Bigl(\delta_{ij} + \overline{h}_{ij}\Bigr)\,dx^i dx^j\,,
\ee
with
\be
\overline{h}_{ij} =(\mathbf{v}_{uli}\,\partial_l \mathcal{I}_{j} + \mathbf{v}_{ulj}\,\partial_l\mathcal{I}_{i} - \delta_{ij}\,\mathbf{v}_{ulk}\,\partial_l \mathcal{I}_{k})\,, 
\ee
where the integral $\mathcal{I}_i$ has been defined in (\ref{integraldef}) and its properties are stated in (\ref{integralprop}).

At first order in $\overline{h}_{ij}$ the curvature of $ds^2_4$ is
\bea
\overline{R}_{ij,kl} &=& \frac{1}{2} \Bigl(\partial_k \partial_j \bar h_{il} - \partial_l \partial_j \bar h_{ik} - (i\leftrightarrow j)\Bigr)\nonumber\\
&\equiv& \Bigl( (1)_{ij,kl} + (2)_{ij,kl} - (3)_{ij,kl} - (i\leftrightarrow j)\Bigr)\,,
\eea
where
\bea
(1)_{ij,kl} &\equiv& \mathbf{v}_{umi}\,\partial_k\,\partial_j\,\partial_m\, \mathcal{I}_{l}- (k\leftrightarrow l)\,,\\
(2)_{ij,kl} &\equiv&  \mathbf{v}_{uml}\,\partial_k\,\partial_j\,\partial_m \,\mathcal{I}_{i} - (k\leftrightarrow l)\,,\\
(3)_{ij,kl} &\equiv& \delta_{il}\,\mathbf{v}_{umn}\,\partial_k\,\partial_j\, \partial_m\, \mathcal{I}_{n}  - (k\leftrightarrow l)\,.
\eea
We want to compute $\frac{1}{2}\,\epsilon_{klpq}\, \overline{R}_{ij,pq}$ and prove that it is equal to $\overline{R}_{ij,kl}$.  We will do it term by term:

(1):
\be
\frac{1}{2}\,\epsilon_{klpq}\,(1)_{ij,pq} =\epsilon_{klpq}\, \mathbf{v}_{umi}\,\partial_p\,\partial_j\, \partial_m\,\mathcal{I}_{q}=-\frac{1}{2}\,\epsilon_{klpq}\, \epsilon_{mirs}\,\mathbf{v}_{urs}\,\partial_p\,\partial_j\,\partial_m\, \mathcal{I}_{q}\,,
\ee
where we have used, in the second equality, the anti-self-duality of $\mathbf{v}_{uij}$. The product of the two epsilon's gives, up to the exchange of $r$ and $s$ that cancels the factor $1/2$, 12 possible terms: the 3 terms proportional to $\delta_{pm}$ and the 3 terms proportional to $\delta_{qm}$ vanish thanks to (\ref{integralprop}); the 2 terms proportional to $\delta_{pi}$ are symmetric in $i,j$ and hence cancel after anti-symmetrization in $i,j$; the 4 terms left give
\bea
\frac{1}{2}\,\epsilon_{klpq}\,(1)_{ij,pq} &=& \Bigl(\mathbf{v}_{upl}\,\partial_p\,\partial_j \,\partial_k\,\mathcal{I}_{i} - (k\leftrightarrow l)\Bigr)-\Bigl(\delta_{il}\,\mathbf{v}_{upq}\,\partial_p\,\partial_j \,\partial_k\,\mathcal{I}_{q}-(k\leftrightarrow l)\Bigr) \nonumber\\
&=& (2)_{ij,kl}-(3)_{ij,kl} \,,
\eea
where the second equality follows from the first property in (\ref{integralprop}).

(2):
\be
\frac{1}{2}\,\epsilon_{klpq}\,(2)_{ij,pq} =\epsilon_{klpq}\,\mathbf{v}_{umq}\,\partial_p\,\partial_j\,\partial_m\, \mathcal{I}_{i} =-\frac{1}{2}\,\epsilon_{klpq}\,\epsilon_{mqrs}\,\mathbf{v}_{urs}\,\partial_p\,\partial_j\, \partial_m\,\mathcal{I}_{i} \,;
\ee
the contraction of the two epsilon's produces, up to $r$ and $s$ exchange,  3 different terms: the term with $\delta_{pm}$ vanishes due to (\ref{integralprop}), and the other 2 terms give
\be
\frac{1}{2}\,\epsilon_{klpq}\,(2)_{ij,pq} = \mathbf{v}_{upl}\,\partial_p\,\partial_j\,\partial_k\,\mathcal{I}_{i}=  (2)_{ij,kl}\,.
\ee

(3):
\be
\frac{1}{2}\,\epsilon_{klpq}\,(3)_{ij,pq} =\epsilon_{klpq}\, \delta_{iq}\,\mathbf{v}_{umn}\,\partial_p\,\partial_j\, \partial_m\,\mathcal{I}_{n}=-\frac{1}{2}\,\epsilon_{klpq}\, \epsilon_{mnrs}\,\delta_{iq}\,\mathbf{v}_{urs}\,\partial_p\,\partial_j\, \partial_m\,\mathcal{I}_{n}\,;
\ee
for the reasons explained above, of the 12 terms coming from the expansion of the two uncontracted epsilon's the ones containing $\delta_{pm}$, $\delta_{im}$ or $\delta_{pn}$ vanish, leaving 
\bea
\frac{1}{2}\,\epsilon_{klpq}\,(3)_{ij,pq} &=& - \Bigl(\mathbf{v}_{upi}\,\partial_p\,\partial_j\,\partial_k\,\mathcal{I}_{l}
-(k \leftrightarrow l)\Bigr) + \Bigl(\mathbf{v}_{upl}\,\partial_p\,\partial_j\,\partial_k\,\mathcal{I}_{i}
-(k \leftrightarrow l)\Bigr)\nonumber\\
&=&-(1)_{ij,kl} + (2)_{ij,kl}\,.
\eea
Putting things together:
\bea
\frac{1}{2}\,\epsilon_{klpq}\, \overline{R}_{ij,pq} &=& \Bigl((2)_{ij,kl}-(3)_{ij,kl}+(2)_{ij,kl} +(1)_{ji,kl}-(2)_{ji,kl}-(i\leftrightarrow j) \Bigr)\nonumber\\
&=&  \overline{R}_{ij,kl} \,,
\eea
which is the identity we wanted to prove.

\providecommand{\href}[2]{#2}\begingroup\raggedright\endgroup


\begin{thebibliography}{10}

\bibitem{Strominger:1996sh}
A.~Strominger and C.~Vafa, ``{Microscopic origin of the Bekenstein-Hawking
  entropy},'' {\em Phys.Lett.} {\bf B379} (1996) 99--104,
  \href{http://arXiv.org/abs/hep-th/9601029}{{\tt hep-th/9601029}}.

\bibitem{Callan:1996dv}
C.~G. Callan and J.~M. Maldacena, ``{D-brane approach to black hole quantum
  mechanics},'' {\em Nucl.Phys.} {\bf B472} (1996) 591--610,
  \href{http://arXiv.org/abs/hep-th/9602043}{{\tt hep-th/9602043}}.

\bibitem{Mathur:2005zp}
S.~D. Mathur, ``{The fuzzball proposal for black holes: An elementary
  review},'' {\em Fortsch. Phys.} {\bf 53} (2005) 793--827,
\href{http://arXiv.org/abs/hep-th/0502050}{{\tt hep-th/0502050}}.
%%CITATION = HEP-TH/0502050;%%.

\bibitem{Mathur:2008nj}
S.~D. Mathur, ``{Fuzzballs and the information paradox: a summary and
  conjectures},''
\href{http://arXiv.org/abs/0810.4525}{{\tt 0810.4525}}.
%%CITATION = 0810.4525;%%.

\bibitem{Skenderis:2008qn}
K.~Skenderis and M.~Taylor, ``{The fuzzball proposal for black holes},'' {\em
  Phys. Rept.} {\bf 467} (2008) 117--171,
\href{http://arXiv.org/abs/0804.0552}{{\tt 0804.0552}}.
%%CITATION = 0804.0552;%%.

\bibitem{Balasubramanian:2008da}
V.~Balasubramanian, J.~de~Boer, S.~El-Showk, and I.~Messamah, ``{Black Holes as
  Effective Geometries},'' {\em Class.Quant.Grav.} {\bf 25} (2008) 214004,
  \href{http://arXiv.org/abs/0811.0263}{{\tt 0811.0263}}.

\bibitem{Chowdhury:2010ct}
B.~D. Chowdhury and A.~Virmani, ``{Modave Lectures on Fuzzballs and Emission
  from the D1-D5 System},''
\href{http://arXiv.org/abs/1001.1444}{{\tt 1001.1444}}.
%%CITATION = 1001.1444;%%.

\bibitem{Hawking:1974sw}
S.~W. Hawking, ``{Particle Creation by Black Holes},'' {\em Commun. Math.
  Phys.} {\bf 43} (1975)
199--220.
%%CITATION = CMPHA,43,199;%%.

\bibitem{Hawking:1976ra}
S.~W. Hawking, ``{Breakdown of Predictability in Gravitational Collapse},''
  {\em Phys. Rev.} {\bf D14} (1976)
2460--2473.
%%CITATION = PHRVA,D14,2460;%%.

\bibitem{Mathur:2009hf}
S.~D. Mathur, ``{The information paradox: A pedagogical introduction},'' {\em
  Class. Quant. Grav.} {\bf 26} (2009) 224001,
\href{http://arXiv.org/abs/0909.1038}{{\tt 0909.1038}}.
%%CITATION = 0909.1038;%%.

\bibitem{Mathur:2011wg}
S.~D. Mathur and C.~J. Plumberg, ``{Correlations in Hawking radiation and the
  infall problem},''
\href{http://arXiv.org/abs/1101.4899}{{\tt 1101.4899}}.
%%CITATION = 1101.4899;%%.

\bibitem{Mathur:2011uj}
S.~D. Mathur, ``{What the information paradox is {\it not}},''
  \href{http://arXiv.org/abs/1108.0302}{{\tt 1108.0302}}.

\bibitem{Bena:2007kg}
I.~Bena and N.~P. Warner, ``{Black holes, black rings and their microstates},''
  {\em Lect. Notes Phys.} {\bf 755} (2008) 1--92,
\href{http://arXiv.org/abs/hep-th/0701216}{{\tt hep-th/0701216}}.
%%CITATION = HEP-TH/0701216;%%.

\bibitem{Dabholkar:1995nc}
A.~Dabholkar, J.~P. Gauntlett, J.~A. Harvey, and D.~Waldram, ``{Strings as
  Solitons \& Black Holes as Strings},'' {\em Nucl. Phys.} {\bf B474} (1996)
  85--121,
\href{http://arXiv.org/abs/hep-th/9511053}{{\tt hep-th/9511053}}.
%%CITATION = HEP-TH/9511053;%%.

\bibitem{Callan:1995hn}
C.~G. Callan, J.~M. Maldacena, and A.~W. Peet, ``{Extremal Black Holes As
  Fundamental Strings},'' {\em Nucl. Phys.} {\bf B475} (1996) 645--678,
\href{http://arXiv.org/abs/hep-th/9510134}{{\tt hep-th/9510134}}.
%%CITATION = HEP-TH/9510134;%%.

\bibitem{Lunin:2001fv}
O.~Lunin and S.~D. Mathur, ``{Metric of the multiply wound rotating string},''
  {\em Nucl. Phys.} {\bf B610} (2001) 49--76,
\href{http://arXiv.org/abs/hep-th/0105136}{{\tt hep-th/0105136}}.
%%CITATION = HEP-TH/0105136;%%.

\bibitem{Lunin:2001jy}
O.~Lunin and S.~D. Mathur, ``{AdS/CFT duality and the black hole information
  paradox},'' {\em Nucl. Phys.} {\bf B623} (2002) 342--394,
\href{http://arXiv.org/abs/hep-th/0109154}{{\tt hep-th/0109154}}.
%%CITATION = HEP-TH/0109154;%%.

\bibitem{Taylor:2005db}
M.~Taylor, ``{General 2 charge geometries},'' {\em JHEP} {\bf 03} (2006) 009,
\href{http://arXiv.org/abs/hep-th/0507223}{{\tt hep-th/0507223}}.
%%CITATION = HEP-TH/0507223;%%.

\bibitem{Kanitscheider:2007wq}
I.~Kanitscheider, K.~Skenderis, and M.~Taylor, ``{Fuzzballs with internal
  excitations},'' {\em JHEP} {\bf 06} (2007) 056,
\href{http://arXiv.org/abs/0704.0690}{{\tt 0704.0690}}.
%%CITATION = 0704.0690;%%.

\bibitem{Bena:2004de}
I.~Bena and N.~P. Warner, ``{One ring to rule them all ... and in the darkness
  bind them?},'' {\em Adv.Theor.Math.Phys.} {\bf 9} (2005) 667--701,
  \href{http://arXiv.org/abs/hep-th/0408106}{{\tt hep-th/0408106}}.

\bibitem{Bena:2005va}
I.~Bena and N.~P. Warner, ``{Bubbling supertubes and foaming black holes},''
  {\em Phys.Rev.} {\bf D74} (2006) 066001,
  \href{http://arXiv.org/abs/hep-th/0505166}{{\tt hep-th/0505166}}.

\bibitem{Berglund:2005vb}
P.~Berglund, E.~G. Gimon, and T.~S. Levi, ``{Supergravity microstates for BPS
  black holes and black rings},'' {\em JHEP} {\bf 0606} (2006) 007,
  \href{http://arXiv.org/abs/hep-th/0505167}{{\tt hep-th/0505167}}.

\bibitem{Bena:2006is}
I.~Bena, C.-W. Wang, and N.~P. Warner, ``{The Foaming three-charge black
  hole},'' {\em Phys.Rev.} {\bf D75} (2007) 124026,
  \href{http://arXiv.org/abs/hep-th/0604110}{{\tt hep-th/0604110}}.

\bibitem{Bena:2007qc}
I.~Bena, C.-W. Wang, and N.~P. Warner, ``{Plumbing the Abyss: Black ring
  microstates},'' {\em JHEP} {\bf 0807} (2008) 019,
  \href{http://arXiv.org/abs/0706.3786}{{\tt 0706.3786}}.

\bibitem{Balasubramanian:2006gi}
V.~Balasubramanian, E.~G. Gimon, and T.~S. Levi, ``{Four Dimensional Black Hole
  Microstates: From D-branes to Spacetime Foam},'' {\em JHEP} {\bf 0801} (2008)
  056, \href{http://arXiv.org/abs/hep-th/0606118}{{\tt hep-th/0606118}}.

\bibitem{deBoer:2008zn}
J.~de~Boer, S.~El-Showk, I.~Messamah, and D.~Van~den Bleeken, ``{Quantizing N=2
  Multicenter Solutions},'' {\em JHEP} {\bf 0905} (2009) 002,
  \href{http://arXiv.org/abs/0807.4556}{{\tt 0807.4556}}.

\bibitem{Bena:2010gg}
I.~Bena, N.~Bobev, S.~Giusto, C.~Ruef, and N.~P. Warner, ``{An
  Infinite-Dimensional Family of Black-Hole Microstate Geometries},'' {\em
  JHEP} {\bf 1103} (2011) 022, \href{http://arXiv.org/abs/1006.3497}{{\tt
  1006.3497}}.

\bibitem{deBoer:2009un}
J.~de~Boer, S.~El-Showk, I.~Messamah, and D.~Van~den Bleeken, ``{A Bound on the
  entropy of supergravity?},'' {\em JHEP} {\bf 1002} (2010) 062,
  \href{http://arXiv.org/abs/0906.0011}{{\tt 0906.0011}}.

\bibitem{Skenderis:2007yb}
K.~Skenderis and M.~Taylor, ``{Anatomy of bubbling solutions},'' {\em JHEP}
  {\bf 0709} (2007) 019, \href{http://arXiv.org/abs/0706.0216}{{\tt
  0706.0216}}.

\bibitem{Giusto:2004id}
S.~Giusto, S.~D. Mathur, and A.~Saxena, ``{Dual geometries for a set of
  3-charge microstates},'' {\em Nucl. Phys.} {\bf B701} (2004) 357--379,
\href{http://arXiv.org/abs/hep-th/0405017}{{\tt hep-th/0405017}}.
%%CITATION = HEP-TH/0405017;%%.

\bibitem{Giusto:2004kj}
S.~Giusto and S.~D. Mathur, ``{Geometry of D1-D5-P bound states},'' {\em Nucl.
  Phys.} {\bf B729} (2005) 203--220,
\href{http://arXiv.org/abs/hep-th/0409067}{{\tt hep-th/0409067}}.
%%CITATION = HEP-TH/0409067;%%.

\bibitem{Ford:2006yb}
J.~Ford, S.~Giusto, and A.~Saxena, ``{A class of BPS time-dependent 3-charge
  microstates from spectral flow},'' {\em Nucl. Phys.} {\bf B790} (2008)
  258--280,
\href{http://arXiv.org/abs/hep-th/0612227}{{\tt hep-th/0612227}}.
%%CITATION = HEP-TH/0612227;%%.

\bibitem{Bena:2011zw}
I.~Bena, B.~D. Chowdhury, J.~de~Boer, S.~El-Showk, and M.~Shigemori,
  ``{Moulting Black Holes},''
\href{http://arXiv.org/abs/1108.0411}{{\tt 1108.0411}}.
%%CITATION = 1108.0411;%%.

\bibitem{DiVecchia:1997pr}
P.~Di~Vecchia {\em et al.}, ``{Classical p-branes from boundary state},'' {\em
  Nucl. Phys.} {\bf B507} (1997) 259--276,
\href{http://arXiv.org/abs/hep-th/9707068}{{\tt hep-th/9707068}}.
%%CITATION = HEP-TH/9707068;%%.

\bibitem{DiVecchia:1999uf}
P.~Di~Vecchia, M.~Frau, A.~Lerda, and A.~Liccardo, ``{(F,Dp) bound states from
  the boundary state},'' {\em Nucl. Phys.} {\bf B565} (2000) 397--426,
\href{http://arXiv.org/abs/hep-th/9906214}{{\tt hep-th/9906214}}.
%%CITATION = HEP-TH/9906214;%%.

\bibitem{Giusto:2009qq}
S.~Giusto, J.~F. Morales, and R.~Russo, ``{D1D5 microstate geometries from
  string amplitudes},'' {\em JHEP} {\bf 03} (2010) 130,
\href{http://arXiv.org/abs/0912.2270}{{\tt 0912.2270}}.
%%CITATION = 0912.2270;%%.

\bibitem{Das:1996ug}
S.~R. Das and S.~D. Mathur, ``{Excitations of D strings, entropy and
  duality},'' {\em Phys.Lett.} {\bf B375} (1996) 103--110,
  \href{http://arXiv.org/abs/hep-th/9601152}{{\tt hep-th/9601152}}.

\bibitem{Hikida:2003bq}
Y.~Hikida, H.~Takayanagi, and T.~Takayanagi, ``{Boundary states for D-branes
  with traveling waves},'' {\em JHEP} {\bf 04} (2003) 032,
\href{http://arXiv.org/abs/hep-th/0303214}{{\tt hep-th/0303214}}.
%%CITATION = HEP-TH/0303214;%%.

\bibitem{Blum:2003if}
J.~D. Blum, ``{Gravitational radiation from travelling waves on D- strings},''
  {\em Phys. Rev.} {\bf D68} (2003) 086003,
\href{http://arXiv.org/abs/hep-th/0304173}{{\tt hep-th/0304173}}.
%%CITATION = HEP-TH/0304173;%%.

\bibitem{Bachas:2003sj}
C.~P. Bachas and M.~R. Gaberdiel, ``{World-sheet duality for D-branes with
  travelling waves},'' {\em JHEP} {\bf 03} (2004) 015,
\href{http://arXiv.org/abs/hep-th/0310017}{{\tt hep-th/0310017}}.
%%CITATION = HEP-TH/0310017;%%.

\bibitem{Black:2010uq}
W.~Black, R.~Russo, and D.~Turton, ``{The supergravity fields for a D-brane
  with a travelling wave from string amplitudes},'' {\em Phys. Lett.} {\bf
  B694} (2010) 246--251,
\href{http://arXiv.org/abs/1007.2856}{{\tt 1007.2856}}.
%%CITATION = 1007.2856;%%.

\bibitem{Gurarie:1993xq}
V.~Gurarie, ``{Logarithmic operators in conformal field theory},'' {\em Nucl.
  Phys.} {\bf B410} (1993) 535--549,
\href{http://arXiv.org/abs/hep-th/9303160}{{\tt hep-th/9303160}}.
%%CITATION = HEP-TH/9303160;%%.

\bibitem{Periwal:1996pw}
V.~Periwal and O.~Tafjord, ``{D-brane recoil},'' {\em Phys. Rev.} {\bf D54}
  (1996) 3690--3692,
\href{http://arXiv.org/abs/hep-th/9603156}{{\tt hep-th/9603156}}.
%%CITATION = HEP-TH/9603156;%%.

\bibitem{Kogan:1996zv}
I.~I. Kogan, N.~E. Mavromatos, and J.~F. Wheater, ``{D-brane recoil and
  logarithmic operators},'' {\em Phys. Lett.} {\bf B387} (1996) 483--491,
\href{http://arXiv.org/abs/hep-th/9606102}{{\tt hep-th/9606102}}.
%%CITATION = HEP-TH/9606102;%%.

\bibitem{Kogan:2000fa}
I.~I. Kogan and J.~F. Wheater, ``{Boundary logarithmic conformal field
  theory},'' {\em Phys. Lett.} {\bf B486} (2000) 353--361,
\href{http://arXiv.org/abs/hep-th/0003184}{{\tt hep-th/0003184}}.
%%CITATION = HEP-TH/0003184;%%.

\bibitem{Lambert:2003zr}
N.~D. Lambert, H.~Liu, and J.~M. Maldacena, ``{Closed strings from decaying
  D-branes},'' {\em JHEP} {\bf 03} (2007) 014,
\href{http://arXiv.org/abs/hep-th/0303139}{{\tt hep-th/0303139}}.
%%CITATION = HEP-TH/0303139;%%.

\bibitem{Bena:2011uw}
I.~Bena, J.~de~Boer, M.~Shigemori, and N.~P. Warner, ``{Double, Double
  Supertube Bubble},''
\href{http://arXiv.org/abs/1107.2650}{{\tt 1107.2650}}.
%%CITATION = 1107.2650;%%.

\bibitem{Balasubramanian:2005mg}
V.~Balasubramanian, J.~de~Boer, V.~Jejjala, and J.~Simon, ``{The library of
  Babel: On the origin of gravitational thermodynamics},'' {\em JHEP} {\bf 12}
  (2005) 006,
\href{http://arXiv.org/abs/hep-th/0508023}{{\tt hep-th/0508023}}.
%%CITATION = HEP-TH/0508023;%%.

\bibitem{Balasubramanian:2006jt}
V.~Balasubramanian, B.~Czech, K.~Larjo, and J.~Simon, ``{Integrability vs.
  information loss: A simple example},'' {\em JHEP} {\bf 11} (2006) 001,
\href{http://arXiv.org/abs/hep-th/0602263}{{\tt hep-th/0602263}}.
%%CITATION = HEP-TH/0602263;%%.

\bibitem{Balasubramanian:2007zt}
V.~Balasubramanian, B.~Czech, K.~Larjo, D.~Marolf, and J.~Simon, ``{Quantum
  geometry and gravitational entropy},'' {\em JHEP} {\bf 12} (2007) 067,
\href{http://arXiv.org/abs/0705.4431}{{\tt 0705.4431}}.
%%CITATION = 0705.4431;%%.

\bibitem{Callan:1988wz}
C.~G. Callan, Jr., C.~Lovelace, C.~R. Nappi, and S.~A. Yost, ``{Loop
  Corrections to Superstring Equations of Motion},'' {\em Nucl. Phys.} {\bf
  B308} (1988)
221.
%%CITATION = NUPHA,B308,221;%%.

\bibitem{Bachas:2002jg}
C.~Bachas, ``{Relativistic string in a pulse},'' {\em Ann. Phys.} {\bf 305}
  (2003) 286--309,
\href{http://arXiv.org/abs/hep-th/0212217}{{\tt hep-th/0212217}}.
%%CITATION = HEP-TH/0212217;%%.

\bibitem{Bertolini:2005qh}
M.~Bertolini, M.~Billo, A.~Lerda, J.~F. Morales, and R.~Russo, ``{Brane world
  effective actions for D-branes with fluxes},'' {\em Nucl.Phys.} {\bf B743}
  (2006) 1--40, \href{http://arXiv.org/abs/hep-th/0512067}{{\tt
  hep-th/0512067}}.

\bibitem{Liu:2002ft}
H.~Liu, G.~W. Moore, and N.~Seiberg, ``{Strings in a time-dependent
  orbifold},'' {\em JHEP} {\bf 06} (2002) 045,
\href{http://arXiv.org/abs/hep-th/0204168}{{\tt hep-th/0204168}}.
%%CITATION = HEP-TH/0204168;%%.

\bibitem{Liu:2002kb}
H.~Liu, G.~W. Moore, and N.~Seiberg, ``{Strings in time-dependent orbifolds},''
  {\em JHEP} {\bf 10} (2002) 031,
\href{http://arXiv.org/abs/hep-th/0206182}{{\tt hep-th/0206182}}.
%%CITATION = HEP-TH/0206182;%%.

\bibitem{Billo:2002hm}
M.~Billo {\em et al.}, ``{Classical gauge instantons from open strings},'' {\em
  JHEP} {\bf 02} (2003) 045,
\href{http://arXiv.org/abs/hep-th/0211250}{{\tt hep-th/0211250}}.
%%CITATION = HEP-TH/0211250;%%.

\bibitem{Billo:2011uc}
M.~Billo, M.~Frau, L.~Giacone, and A.~Lerda, ``{Holographic non-perturbative
  corrections to gauge couplings},'' {\em JHEP} {\bf 1108} (2011) 007,
  \href{http://arXiv.org/abs/1105.1869}{{\tt 1105.1869}}.

\bibitem{Polchinski:1998rr}
J.~Polchinski, {\em {String theory. Vol. 2: Superstring Theory and Beyond}}.
\newblock Cambridge Univ. Pr., 1998.

\bibitem{Bergman:1994qq}
O.~Bergman and B.~Zwiebach, ``{The Dilaton theorem and closed string
  backgrounds},'' {\em Nucl. Phys.} {\bf B441} (1995) 76--118,
\href{http://arXiv.org/abs/hep-th/9411047}{{\tt hep-th/9411047}}.
%%CITATION = HEP-TH/9411047;%%.

\bibitem{Billo:1998vr}
M.~Billo {\em et al.}, ``{Microscopic string analysis of the D0-D8 brane system
  and dual R-R states},'' {\em Nucl. Phys.} {\bf B526} (1998) 199--228,
\href{http://arXiv.org/abs/hep-th/9802088}{{\tt hep-th/9802088}}.
%%CITATION = HEP-TH/9802088;%%.

\end{thebibliography}
\end{document}